\documentclass[twocolumn, twocolappendix, dvipsnames]{aastex63}

\submitjournal{ApJ}

\usepackage{amsmath,natbib, float, color, soul, xspace, tabularx,dsfont, cancel}
\usepackage[normalem]{ulem}
\usepackage{listings}

\usepackage{array}
\newcolumntype{R}[1]{>{\raggedright\arraybackslash\hspace{0pt}}p{#1}}

\newcommand*\msolarh[0]{h^{-1} \, \mathrm{M_{\odot}}}

\newcommand*\sig{\textsc{$\sigma_8$}\xspace}

\newcommand*\Mgas{\textsc{$M_\mathrm{gas}$}\xspace}

\newcommand*\chandra{\textit{Chandra}\xspace}
\newcommand*\mgas{\textsc{$M_\mathrm{gas}$}\xspace}
\newcommand*\mpch{\textsc{$h^{-1}\,\mathrm{Mpc}$}\xspace}
\newcommand*\tng{\texttt{IllustrisTNG}\xspace}
\newcommand*\magneticum{\texttt{Magneticum}\xspace}

\newcommand*\yx{$Y_X$\xspace}
\newcommand*\multimodel{Scatter-Augmented\xspace}

\shorttitle{Domain Adaptation Cluster Mass Proxy}
\shortauthors{Ntampaka et al.}

\begin{document}

\correspondingauthor{Michelle Ntampaka}
\email{mntampaka@stsci.edu}

\author[0000-0002-0144-387X]{Michelle Ntampaka}
\affiliation{Space Telescope Science Institute, Baltimore, MD 21218, USA}
\affiliation{Department of Physics \& Astronomy, Johns Hopkins University, Baltimore, MD 21218, USA}

\author[0000-0003-1281-7192]{A.~\'Ciprijanovi\'c}
\affiliation{Fermi National Accelerator Laboratory, Batavia, IL 60510}
\affiliation{Department of Astronomy and Astrophysics, University of Chicago, Chicago, IL 60637}
\affiliation{NSF-Simons AI Institute for the Sky (SkAI), Chicago, IL 60611, USA}

\author[0009-0000-2927-2104]{Ana Maria Delgado}
\affiliation{Department of Physics \& Astronomy, Johns Hopkins University, Baltimore, MD 21218, USA}

\author[0000-0002-0104-3593]{John Soltis}
\affiliation{Department of Physics \& Astronomy, Johns Hopkins University, Baltimore, MD 21218, USA}

\author[0000-0002-5077-881X]{John F. Wu}
\affiliation{Space Telescope Science Institute, Baltimore, MD 21218, USA}
\affiliation{Department of Physics \& Astronomy, Johns Hopkins University, Baltimore, MD 21218, USA}
\affiliation{Department of Computer Science, Johns Hopkins University, Baltimore, MD 21218, USA}

\author[0000-0002-9851-2850]{Mikaeel Yunus}
\affiliation{Department of Physics \& Astronomy, Johns Hopkins University, Baltimore, MD 21218, USA}

\author[0000-0003-3175-2347]{John ZuHone}
\affiliation{Center for Astrophysics $|$ Harvard \& Smithsonian, Cambridge, MA 02138, USA}

\title{The Importance of Being Adaptable:\\ An Exploration of the Power and Limitations of Domain Adaptation\\for Simulation-Based Inference with Galaxy Clusters} 

\begin{abstract}
    The application of deep machine learning methods in astronomy has exploded in the last decade, with new models showing remarkably improved precision against statistical methods for benchmark tasks.  Not nearly enough attention, however, is given to understanding the models' robustness, especially when the Test Data are systematically different from the Training Data, or ``out of domain.''  Domain shift poses a significant challenge for simulation-based inference, where models are trained on simulated data but applied to real observational data.  
    In this paper, we explore domain shift and test domain adaptation methods for a specific scientific case: simulation-based inference for estimating galaxy cluster masses from X-ray profiles.  We build the following data sets to mimic simulation-based inference: a Training Set from the \magneticum{} simulation, a second Training Set that is scatter-augmented to capture uncertainties in scaling relations, and a Test Set derived from the \tng simulation.  We demonstrate that the Test Set is out of domain in subtle ways that, in a realistic scenario, would be difficult to detect without careful analysis.  We apply three deep learning methods:  a standard and uncorrected neural network (NN), a neural network trained on the scatter-augmented input catalogs, and a Deep Reconstruction-Regression Network (DRRN), a semi-supervised deep model engineered to address domain shift.  Though the NN improves results by a factor of $17\%$ in the Training Data, it performs $40\%$ worse on the out-of-domain Test Set.  Surprisingly, the \multimodel Neural Network (SANN) performs similarly.  While the DRRN is successful in mapping the training and Test Data onto the same latent space, it consistently underperforms compared to a straightforward \yx scaling relation.  These results serve as a warning that simulation-based inference must be handled with extreme care, as subtle differences between training simulations and observational data can lead to unforeseen biases creeping into the results.
	\vspace{12ex}
\end{abstract}


\section*{}\bigskip

\section{Introduction}
\label{sec:intro}

The modern cosmological model, $\Lambda$CDM, describes a Universe with structure formation driven by cold dark matter (CDM) and undergoing accelerated expansion due to a cosmological constant ($\Lambda$).  While this model has incredible predictive power to describe the 3D distribution of matter, there remain unresolved tensions in the model, especially between early- and late-time measurements \citep[see measurements and discussions of this tension in, e.g.,][]{2016A&A...594A..24P, 2019ApJ...880..154N, 2021A&A...646A.140H, 2022PhRvD.106d3520P, 2024A&A...689A.298G, 2025JCAP...02..045K}.  Recent efforts to measure cosmological parameters at the percent level are not simply an effort to tamp down error bars, but are driven in part by the goal of resolving these tensions to understand whether they stem from systematic errors that have been unaccounted for, or whether they represent deviations from our physical model of the Universe.

Cosmological parameter estimation can be approached using a variety of observables and techniques, including the abundance of galaxy clusters \citep[e.g.,][]{1998ApJ...504....1B, 2009ApJ...692.1060V, 2024PhRvD.110h3510B}, the power spectrum of 3D matter \citep[e.g.,][]{2004ApJ...606..702T}, the temperature and polarization anisotropies of the cosmic microwave background \citep[e.g.,][]{2020A&A...641A...6P}, the distance-redshift relation probed by type Ia supernovae \citep[e.g.,][]{1998AJ....116.1009R, 2022ApJ...938..110B}, shape distortions of background galaxies due to weak lensing \citep[e.g.,][]{1992ApJ...388..272K}, and by probing cosmic distances using baryon acoustic oscillations \citep[e.g.,][]{2005ApJ...633..560E}. 

Machine learning (ML) is a tempting tool to apply to any of these approaches.  ML often boasts the ability to significantly reduce error bars, and early proof-of-concept methods for estimating cosmological parameters with ML have shown promise \citep[e.g.,][]{2017arXiv170705167S, 2018PhRvD..97j3515G, 2020ApJ...889..151N}. Applying an ML-based method to estimate cosmological parameters directly from observations typically relies on ``simulation-based inference,\footnote{Throughout this work, we use the most generic definition of ``simulation-based inference'' to describe the use of any deep learning model for training on simulations and applying to observed data.}'' an approach in which the ML model is trained on simulations and the model is applied to observational data \citep[e.g.,][]{2024arXiv240404228M, 2024arXiv240920507R, 2025arXiv250313755B, 2025PhRvD.111h3510N}. 

Unfortunately, machine learning and especially deep learning methods often exhibit a substantial drop in performance when simulation-trained models are applied to real data\textemdash{} an effect called the domain shift problem. These differences can be caused by imperfect simulations (due to unknown physics or computational constraints), as well as our inability to perfectly mimic observational effects such as noise, PSF blurring of the telescope, or potential problems with the detector itself. Domain adaptation methods~\citep{Wang_2018,wilsongarrett} enable deep learning models to find and utilize only domain-invariant features (i.e., features present in both simulated and observational data). In recent years, domain adaptation has enabled the creation of robust deep learning models for different astronomical applications that require the model to work across multiple datasets, for example:  galaxy morphology classification~\citep{deepadversaries,sidda},  constraining star formation histories of galaxies~\citep{Gilda2024}, gravitational lens finding~\citep{Parul2024}, inference of strong gravitational lens parameters~\citep{Swierc2023,Agarwal2024}, and inference of cosmology~\citep{Roncoli2023}.  

When training on simulated data with the prospect of applying the model to real data, correctly quantifying and mitigating the sources of bias is extremely challenging \citep[see, for example, the discussion in][]{2023MLS&T...4aLT01L}.  
Hence, bias mitigation and deeper understanding of the role of calibration must first be addressed in a simpler regime, where methods can be rigorously tested, uncertainties are robust and understood, and\textemdash{}critically\textemdash{}the results can be trusted. This paper examines bias in one such simpler regime, where traditional techniques for quantifying and correcting bias \textit{are} available (in the form of calibration).  However, the methods we explore and the limitations we describe in this manuscript are broadly applicable beyond the narrow scientific case we explore.  The simpler regime we will explore in this paper models simulation-based inference for galaxy cluster mass estimation.

Galaxy clusters are massive, gravitationally bound systems.  They contain hundreds to thousands of galaxies and a hot intracluster medium (ICM), all of which are embedded in a massive dark matter halo.  These massive, gravitationally bound objects represent the high-mass tail of the halo mass function, and their abundance is sensitive to the underlying cosmological parameters. 

Using cluster abundance for cosmological analysis requires a sample of clusters with a well-understood selection function and a method for interpreting cluster observations in terms of cluster mass.  
Optical mass proxies such as richness \citep[e.g.,][]{2007arXiv0709.1159J}, velocity dispersion \citep{zwicky1933rotverschiebung, 2016ApJ...819...63R}, and weak lensing \citep{2014MNRAS.439...48A} probe the galaxies in, around, or behind the cluster, while the Sunyaev-Zeldovich (SZ) Effect in the microwave \citep[e.g.,][]{1972CoASP...4..173S, 2015ApJS..216...27B} and thermal bremsstrahlung radiation plus collisionally excited line emission in the X-ray \citep{2006ApJ...640..691V} are mass proxies that utilize the intra-cluster medium, a hot plasma sitting in the cluster's gravitational potential.  

These gas-based mass proxies (SZ and X-ray) are high-precision, low-accuracy mass proxies, meaning that they give low-scatter results that are likely to be biased and need to be calibrated. This bias is believed to be primarily caused by non-thermal pressure support, sources of pressure in the ICM that arise from sources other than the thermal motion of the particles \citep[e.g.,][]{Lau:2009aa, 2012ApJ...751..121N, 2014ApJ...792...25N, 2014MNRAS.442..521S, 2015MNRAS.448.1020S, 2019A&A...621A..40E,
2025arXiv250506533X}, including bulk motions, turbulence, or magnetic fields.  Uncertainties in gas physics make it particularly challenging to model the ICM in a way that is robust across scales ranging from galaxy morphologies to cluster density profiles in cosmological hydrodynamical simulations.

Weak lensing provides a direct probe of the total mass distribution with relatively low bias but suffers from significant scatter on a per-object basis. In contrast, X-ray and SZ mass proxies, including the X-ray observable \yx \citep{2006ApJ...650..128K}, exhibit low intrinsic scatter, but are susceptible to systematic biases arising from uncertainties in gas physics and non-thermal pressure support. To capitalize on the low scatter of these observables while mitigating their biases, they are often calibrated against weak lensing masses. This approach uses weak lensing as a high-accuracy anchor for the normalization of the mass–observable relation, and it uses X-ray or SZ proxies for high-precision relative mass estimates within a cluster sample.

Because uncertainties in gas physics and non-thermal pressure support are significant sources of errors in X-ray-based observations of clusters, because they are difficult to model, and because they introduce a space to test both precision and accuracy, gas-based observations are an interesting place to develop and test deep learning models that may likewise be susceptible to bias.  

In Section \ref{sec:mocks}, we describe the mock observational data sets that have been developed with very different underlying assumptions affecting the X-ray profiles of the clusters; these two data sets are described in detail in Tables \ref{table:sim_compare} and \ref{table:obs_compare}.  The mock observations are intentionally different to make a realistic test of simulation-based inference by training on one simulation and testing on another.

In Section \ref{sec:MLmethods}, we describe several deep learning methods to estimate cluster masses, with varying levels of flexibility built into the data and models to address domain adaptation.  We discuss the positive and negative results, as well as their implications for simulation-based cosmological inference, in Section \ref{sec:results}, and we conclude in Section \ref{sec:conclusions}. \\

\section{Methods: Simulated Galaxy Cluster Observations}
\label{sec:mocks}

\begin{figure*}[t!]
	\centering
	\includegraphics[width=0.9\textwidth]{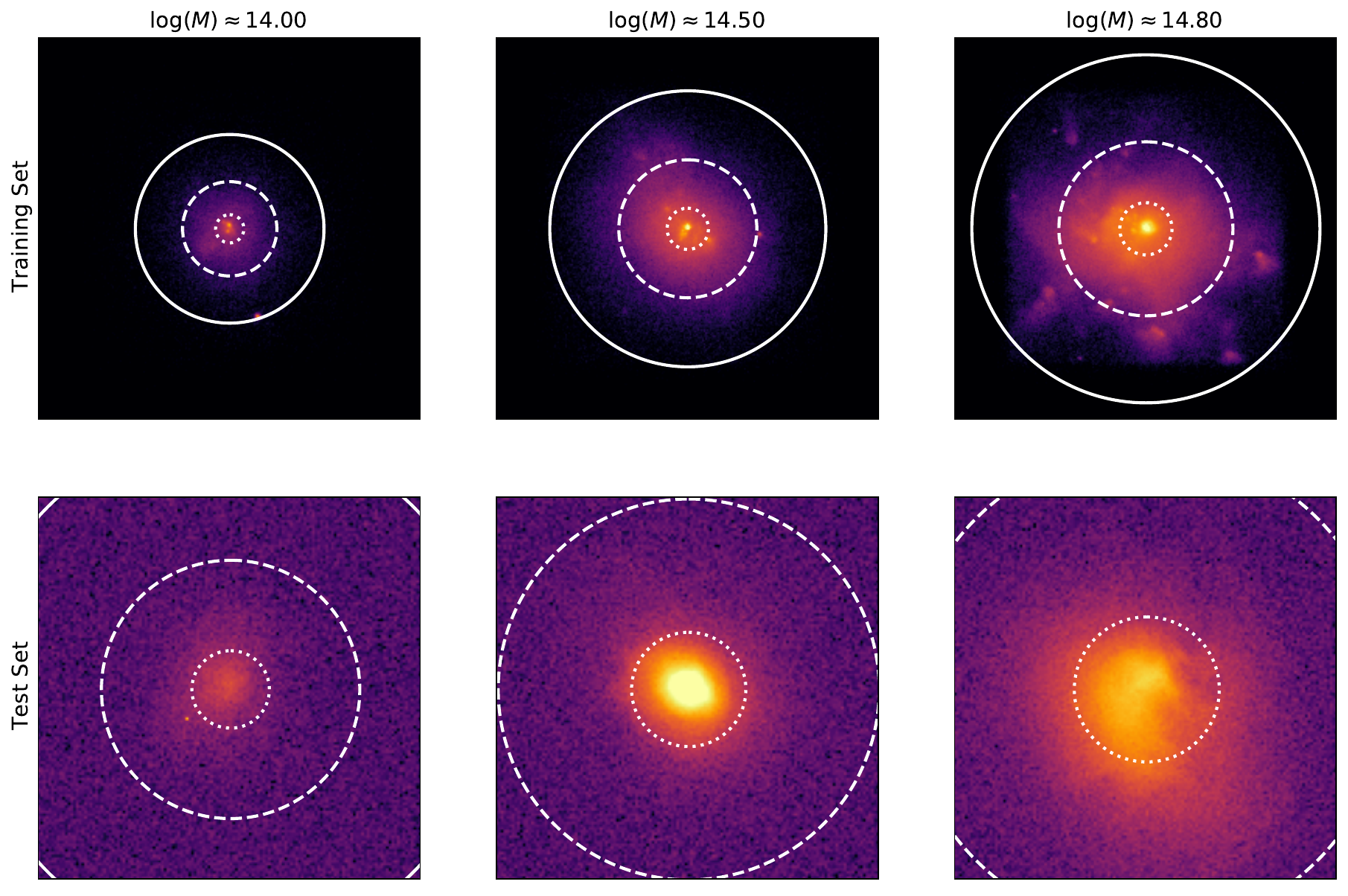} 
		\caption[]{Sample clusters from the Training Set (top row) and Test Set (bottom row) at three masses ($\log\left[M/(h^{-1}M_\odot\right]\approx 14.00, 14.50, 14.80$).  The Training Set is derived from the \texttt{Magneticum} simulation, at z=0.07, as observed for 100ks by a generic instrument with a flat instrument response.  In contrast, the Test Set is derived from \texttt{IllustrisTNG}300 at z=0.05 with a simulated 100ks \chandra{} observation.  Circles denote 0.15$R_\mathrm{500}$ (dotted), 0.50$R_\mathrm{500}$ (dashed), and 1.0$R_\mathrm{500}$ (solid).  The differences between the Training and Test Sets are described in Tables \ref{table:sim_compare} and \ref{table:obs_compare}.  Figures \ref{fig:sully}, \ref{fig:powerlaw}, and \ref{fig:yx2} illustrate further differences between these two data sets.  The Training and Test Sets are visibly different and were constructed this way by design \textemdash{} because ML models can unfairly use simulation artifacts to infer underlying parameters, we built a completely unique Test Set to evaluate the robustness of our model to domain shift.}
       	\label{fig:clustersampleimages}	
\end{figure*}

Supervised machine learning methods rely on labeled data for training. A classical approach is to learn from inexpensive data to output expensive results \citep[for example,][learned to predict high-resolution images from low-resolution inputs]{2022ApJ...940...60S} or to learn from observational data to predict unseen underlying physical properties \citep[as in, for example,][]{2015ApJ...803...50N}.  
In our case, the output is cluster mass, which is not a value that can definitively be extracted from any observations; therefore, we rely on simulations to provide Training Data and the associated labels.

One potential failure mode of using simulated data is that the model will ``overspecialize,'' meaning that the model learns from unphysical artifacts in simulated data and, therefore, may not be reliably applied to observed data\footnote{See Appendix B in \cite{2022ApJ...926...45N} for a discussion of overspecialization and an a case study demonstrating ML ``cheating'' by learning from nonphysical simulation artifacts.}.  The domain shift from simulated to observed data can introduce catastrophic failures in model outputs \citep[e.g.,][]{2021MNRAS.506..677C,2024A&A...687A..24M}.  This is a different failure mode than that of ``overfitting,'' in which the model learns uninformative or non-physical patterns in the Training Data.  In the case of overfitting, the model will fail on the Test Set and will give pessimistic results.  In the case of overspecialization, the model will perform extremely well on any Test Sets derived from the same simulation but will fail on test sets with different properties, such as those with different origins.  

To explore a realistic domain adaptation scenario, we build several data sets that will serve as a proxy simulation-based inference: the Training and Validation Sets (described in \S\ref{sec:trainingset}) are created from the \magneticum simulations, the \multimodel Training Set (also described in \S\ref{sec:trainingset}) is built from \magneticum but is perturbed to capture uncertainties in physical models, and the Test Set (described in \S\ref{sec:testingset}) is created from the \tng simulation.  We compare and contrast these in \S\ref{sec:compare} and in Tables \ref{table:sim_compare} and \ref{table:obs_compare} to show the ways in which the Test Set is out of domain compared to the Training Data.

\subsection{Training and Validation Sets}
\label{sec:trainingset}

\begin{table*}[t] 
\caption{Comparison of Training, Validation, and Test Sets: Simulations}
\label{table:sim_compare}
\begin{center}
\begin{tabular}{| R{4cm} |  R{3.5cm} | R{3.5cm} | R{5cm}  |}
\hline
\hline
Property & Training and Validation  & Test & Comments  \\
\hline
\hline
	\textbf{Simulation}
	& \texttt{Magneticum} Box2
	& \texttt{IllustrisTNG}300
	& See \cite{simcompare} for an overview comparison of these and other recent hydrodynamic simulations.\\
\hline
	Box Length
	& $352\,\mpch$ 
	& $205\,\mpch$ 
	& \\
\hline
	Box Volume
	& $0.0436\,h^{-3}\,\mathrm{Gpc}^3$ 
	& $0.0086\,h^{-3}\,\mathrm{Gpc}^3$
	& \\
\hline
	$N_\mathrm{gas}$
	& $1584^3$
	& $2500^3$ 
	& Number of gas particles.\\
\hline
	$N_\mathrm{DM}$
	&$1584^3$
	& $2500^3$ 
	& Number of dark matter particles.\\
\hline
	$m_\mathrm{gas}$
	& $1.4\times10^{8}\,\msolarh$
	& $7.6\times10^{6}\,\msolarh$ 
	& Mass of gas particles.\\
\hline
	$m_\mathrm{DM}$
	& $6.9\times10^{8}\,\msolarh$ 
	& $4.0\times10^{7}\,\msolarh$ 
	& Mass of dark matter particles.\\

\hline
	$\epsilon$
	&$3.75h^{-1}\,\mathrm{kpc}$
	& $1h^{-1}\,\mathrm{kpc}$ 
	& Gravitational softening length.\\
\hline
	Code Base
	&  \texttt{GADGET3}, a smooth particle hydrodynamics (SPH) approach \mbox{\citep{2008MNRAS.391.1685S}}.
	&  \texttt{AREPO}, a moving mesh approach  \mbox{\citep{sph_weirdkey}}.
	& \\
\hline
Stellar Formation and Evolution
	&  Used the model from \cite{2003MNRAS.339..289S}.
	&  Used the model from \cite{2003MNRAS.339..289S}.
	& The implemented models in each simulation employed different characteristic density and timescales; see simulation papers for details.\\
\hline
	Stellar Feedback
	& Wind particles stochastically spawned from star-forming gas particles; see \cite{2014MNRAS.444.2938H} for details.
	& Wind particles stochastically spawned from star-forming gas cells; see \cite{2018MNRAS.473.4077P} for details.
	& \\
\hline
	Super Massive Black Hole (SMBH) Feedback
	& Two modes at low and high mass accretion rates.  In both modes, continuous thermal dump, but more energetic at low-accretion states
	& Two modes at low and high mass accretion rates.  Continuous thermal dump at high-accretion states; isotropic and pulsated SMBH-driven winds at low-accretion states.
	& The simulations use different choices for the definition of low vs.~high state.\\
\hline
	For additional simulation details, see\ldots
	&  \cite{2016MNRAS.463.1797D}
	&  \cite{2015AandC....13...12N}
	& \\
\hline
\hline
	\textbf{Cosmology}
	& Consistent with WMAP7 \citep{2011ApJS..192...18K}
	& Consistent with \cite{2015arXiv150201589P}
	& \\
\hline	
	$\mathcal{S}_8$
	& 0.789 (low $\mathcal{S}_8$)
	& 0.822 (high $\mathcal{S}_8$)
	& $\mathcal{S}_8\equiv \sigma_8 \left( \frac{\Omega_m}{0.3}\right)^{0.25}$ \\
\hline	
	$\sigma_8$
	& 0.809
	& 0.816
	& \\
\hline
	$\Omega_m$
	& 0.272
	& 0.309
	& \\
\hline
	$\Omega_b$
	& 0.046
	& 0.049
	& \\
\hline
	$\Omega_\Lambda$
	& 0.728
	& 0.691
	& \\
\hline
	$h$
	& 0.704
	& 0.677
	& \\
\hline
	$n_s$
	& 0.963 
	& 0.967
	& \\
\hline
\hline
\end{tabular}
\end{center}
\end{table*}
\begin{table*}[t] 
\caption{Comparison of Standard Training, \multimodel Training, Validation, and Test Sets: Mock Observations \& Cluster Sample}
\label{table:obs_compare}
\begin{center}
\begin{tabular}{| R{4cm} |  R{3.5cm} | R{3.5cm} | R{5cm}  |}
\hline
\hline
Property & Training and Validation  & Test & Comments  \\
\hline
\hline
	\textbf{Simulation}
	& \texttt{Magneticum} Box2
	& \texttt{IllustrisTNG}300
	& See Table \ref{table:sim_compare} for a comparison of these simulations.\\
\hline
\hline
	\textbf{Mock Observations}
	& see \S\ref{sec:trainingset}
	& see \S\ref{sec:testingset}
	& \\
\hline
	Observing Instrument
	& Generic instrument with an ideal (flat) instrument response.
	& \textit{Chandra} field of view and instrument response, but with a lower spatial resolution.
	&  \\
\hline
	Observing Time
	& $100$ks
	& $100$ks
	& \\
\hline
	Halo Finder
	& Friends-of-Friends
	& Friends-of-Friends
	& \\
\hline
	Cluster Redshift
	& $z=0.07$
	& $z=0.05$
	& \\
\hline

	Mock Observation Method
	& \texttt{PHOX}
	& \texttt{pyXSIM}
	& \texttt{pyXSIM} is an implementation of the \texttt{PHOX} algorithm.\\
\hline
	Mock observations originally described in\ldots
	& \cite{2022ApJ...926...45N}
	& \cite{2019ApJ...876...82N}
	& \\
\hline
\hline
	\textbf{Cluster Sample}
	& 
	& 
	& \\
\hline
	$N_\mathrm{cl}$
	& 1148
	& 329
	& Number of unique clusters in the data set.\\
\hline
	$N_\mathrm{obs}$
	& 1148
	& 986
	& Number of cluster observations; in the test set, clusters are viewed from up to 3 lines of sight.\\
\hline
	$M_\mathrm{min}$
	& $10^{13.6} \msolarh$
	& $10^{13.4} \msolarh$
	&  Minimum cluster  $\log(M_{500})$. \\
\hline
\hline
	\textbf{Cluster Power Law Fit Parameters}
	& 
	& 
	& $\sigma$ denotes scatter in mass (in dex) that results from applying the power law.  Fitting the Testing Set with a more flexible broken power law \citep[as motivated by][]{2022arXiv220511528P} does not result in significant changes to the errors listed.\\
\hline
	$M_\mathrm{500c}$-$M_\mathrm{gas}$ fit parameters
	& $\mathcal{A} = -4.55$ \vfill 
		$\alpha  =1.25$ \vfill 
		$\sigma  = 0.05$
	& $\mathcal{A} = -4.00$ \vfill 
		$\alpha  =1.21$ \vfill 
		$\sigma  = 0.03$
	& See \S\ref{sec:mgasT} \\
\hline	
	$M_\mathrm{500c}$-$T$ fit parameters
	& $\mathcal{A} = -8.34$ \vfill 
		$\alpha  = 0.62$\vfill 
		$\sigma  = 0.05$
	& $\mathcal{A} = -7.61$ \vfill 
		$\alpha  =0.56$ \vfill 
		$\sigma  = 0.04$
	&  See \S\ref{sec:mgasT}\\
\hline	
	$M_\mathrm{500c}$-$Y_X$ fit parameters
	& $\mathcal{A}  = -12.90$ \vfill 
		$\alpha  = 1.87$\vfill 
		$\sigma  = 0.03$
	& $\mathcal{A}  = -11.52$ \vfill 
		$\alpha  = 1.76$\vfill 
		$\sigma  = 0.02$
	&   See \S\ref{sec:yx}\\ 
\hline
\hline
\textbf{Derived Catalogs}
	&  \textbf{Standard Training Set}:  90\% of the unique clusters.\vfill
           \textbf{\multimodel Set}:  flat HMF with 300 clusters per 0.1 dex mass bin.\vfill
           \textbf{Validation Set}:  10\% of the unique clusters.
           
	&  \textbf{Test Set}
	& \\
\hline
\hline
\end{tabular}
\end{center}
\end{table*}

The \magneticum Box2 Simulation \citep{2015ApJ...812...29T, 2016MNRAS.463.1797D, 2016MNRAS.456.2361B, 2017MNRAS.469..787P, 2017Galax...5...49R} is a cosmological volume hydrodynamical simulation.  It is selected as the basis for the Training and Validation Sets because the large volume is sufficient to yield a large number of unique massive galaxy clusters for training the model, while the high resolution models the internal structure of each cluster and captures, for example, the clumpy structure that results from recent merger events.  The simulation assumes a low-\sig cosmology consistent with WMAP7 constraints \citep{2011ApJS..192...18K}. Details about the box size, particle resolution, subgrid physics modeling (including stellar formation and evolution, stellar feedback modeling, and supermassive black hole feedback), and cosmology are summarized in Table \ref{table:sim_compare} and detailed in \cite{2016MNRAS.463.1797D}. 

We make the simplest mass and redshift cuts for the Training and Validation Sets, probing the regime where clusters have sufficient X-ray luminosity to infer a robust gas mass profile, and selecting one low redshift snapshot, akin to working under the assumption that cluster redshift is known.  All clusters at $z=0.07$ with a masses  $M_\mathrm{500c} \geq 10^{13.6}\,h^{-1}\,M_\odot$ are selected to comprise the Training and Validation Sets; there are 1148 such clusters in the \magneticum Box2 simulation.  Each cluster is observed from a single line of sight.

We create X-ray mock observations for each cluster using the \magneticum{} online simulator.  Our mock observations were generated for a generic instrument with a flat effective area, $A=600\,\mathrm{cm}^2$ in the 0.5-2.0keV energy band, a square $42'\times42'$ field of view, $6''$ pixels, and with an idealized assumption of no background and infinitely sharp resolution.  We model a 100ks X-ray observation with the \texttt{PHOX} software package \citep{2012MNRAS.420.3545B}.  Three sample X-ray images are shown in the top panel of Figure \ref{fig:clustersampleimages}.  Table \ref{table:obs_compare} summarizes these details about the Training Set mock observations, which are described in further detail in \cite{2022ApJ...926...45N}.

Observed gas mass (from here forward, simply ``gas mass,'' $M_\mathrm{gas}$) is not simply extracted from simulation catalogs.  Instead, we take an approach that more accurately recreates an observational approach and infer gas mass profiles from each mock observation by modeling and fitting the spherically averaged luminosity function, deprojecting this, and inferring the underlying spherically averaged gas mass.  In addition to the gas masses, the mass-weighted temperature ($T$) as given by the \magneticum cluster catalog are also used as cluster inputs for this analysis.  These processes are described in more detail in \S2.1.2 of \cite{2022ApJ...926...45N}.  

Each cluster is assigned to be in either the Training or the Validation Set, using a random 90/10 split.

The Training Set is also used to create a second catalog, the \multimodel Training Set, which is a data set used for training that is more optimized for a domain-adaptive approach.  The \multimodel Training Set differs from the Training Set in two key ways:  scatter is introduced in power law scaling relations, and it has a flat mass function.

Following \cite{2022ApJ...926...45N}, we introduce sources of scatter to build a \multimodel Training Set that marginalizes over uncertainties in the $M-T$ and $M-M_\mathrm{gas}$ scaling relations.  In a simplified version of what is found in \cite{2022ApJ...926...45N}, we create additional copies of each cluster which randomly move $T$ and $M_\mathrm{gas}$ up to a factor of $2\sigma$ closer to or farther from the best fit scaling relation.  We scale the gas mass profiles to match the new $M_\mathrm{gas}$ value.  Thus ``puffing out'' the way the clusters' filling of the $M-T-M_\mathrm{gas}$ space while retaining correlations among physical properties.  This is done in an effort to capture the full uncertainty in these scaling relations, with the goal of having Test Set in the domain of the \multimodel Training Set.

To provide additional Training Data for rare, high-mass objects, the \multimodel Training Set has a flat mass function, achieved by randomly downsampling this catalog to include only 300 clusters per 0.1 dex mass bin.  Due to the rarity of very high-mass clusters and the resulting small number of total clusters in high mass bins, some 0.1 dex bins have no clusters to contribute to the \multimodel catalog.  

To summarize:  the Training Set contains 90\% of the clusters in the \magneticum simulation, for a total of 1041 unique clusters (throughout this paper, when extra clarity is needed, this will be referred to as the ``Standard Training Set'').  The \multimodel Training Set is built from perturbed copies of these same clusters.  It contains 4200 clusters with a flat mass function, and it captures significant uncertainties in scaling relations.  While the Training Set and the \multimodel  Training Set use the same clusters, the clusters in the Validation Set are a completely unique set of 107 clusters randomly selected from the \magneticum simulation; for training, all models use this Validation Set (with no perturbations).  The Standard and \multimodel Training Sets are used to train models, and the Validation Set is used to identify when the models are finished training (look ahead to Section \ref{sec:MLmethods} for details about model training).  Tables \ref{table:sim_compare} and \ref{table:obs_compare} summarize the parameters of the Training, \multimodel, and Validation Sets.

\subsection{Test Set}
\label{sec:testingset}

We create a Test Set that is, by design, still realistic but likely to be out-of-domain compared to the Training Set.  By selecting a simulation with different subgrid physics models (and by utilizing a different mock observation scheme applied to that simulation), we manufacture a scenario that closely mimics real simulation-based inference:  subtle differences between the Training and Test Sets might not manifest in the simplest checks, but are likely to bias the outputs of deep learning.

The \tng 300 Simulation \citep{2018MNRAS.475..648P, 2018MNRAS.475..676S, 2018MNRAS.475..624N, 2018MNRAS.477.1206N, 2018MNRAS.480.5113M, nelson2021illustristng} was selected as the basis for the Testing Set because it makes different subgrid model choices, making it likely to be out-of-domain from the Training Data.  Though the \tng 300 simulation is a significantly smaller volume than \magneticum, it is sufficiently large to produce a Test Set of galaxy clusters for assessing the ability of our model to adapt to out-of-domain data.  

While \magneticum assumes a low-\sig cosmology, this simulation assumes a high-\sig cosmology consistent with Planck cosmological constraints \citep{2018arXiv180706209P}. Further details about the box size, particle resolution, subgrid physics modeling (including stellar formation and evolution, stellar feedback modeling, and supermassive black hole feedback), and cosmology are discussed in \cite{2015AandC....13...12N}.  These parameters are summarized and contrasted with the \magneticum simulation in Table \ref{table:sim_compare}.

Clusters at $z=0.00$ with a masses  $M_\mathrm{500c} \geq 10^{13.4}\,h^{-1}\,M_\odot$ are selected to construct the testing sample of galaxy clusters; there are 329 such clusters in the \tng 300 simulation.  Each cluster is observed from three lines of sight, yielding 986 cluster mock observations.

As with the Training Set data, X-ray mock observations are made for each cluster, but we use more realistic instrument and noise models for this set.  We first assume that each cluster is at a distance of $z=0.05$ from the observer and model a \chandra{} field of view and instrument response for a 100ks X-ray observation using the \texttt{pyXSIM} software package \citep{zuhone-proc-scipy-2014}. The bottom panel of Figure \ref{fig:clustersampleimages} shows these mock X-ray images from a sample of three clusters in the Testing Set that have similar masses to the counterparts shown from \magneticum in the top panel.  Table \ref{table:obs_compare} summarizes these details about the Testing Set mock observations and contrasts them to the Training Set.  This data set is described in further detail in \cite{2019ApJ...876...82N}.

Gas mass ($M_\mathrm{gas}$) is inferred from each mock observation in a process identical to the one described in \S\ref{sec:trainingset} and cluster temperatures ($T$) are taken directly from the simulation catalogs. Unlike the process for the Training Sample, we only use the simulation output clusters for the testing sample.  This is because in \S\ref{sec:trainingset}, we are creating a population of clusters that allows us to marginalize over many possible models.  For the testing set, however, we choose to just use the single simulation output for evaluation.  

\subsection{Comparison of Training and Testing Sets}
\label{sec:compare}

\begin{figure*}[h]
	\centering
	\includegraphics[width=0.9 \textwidth]{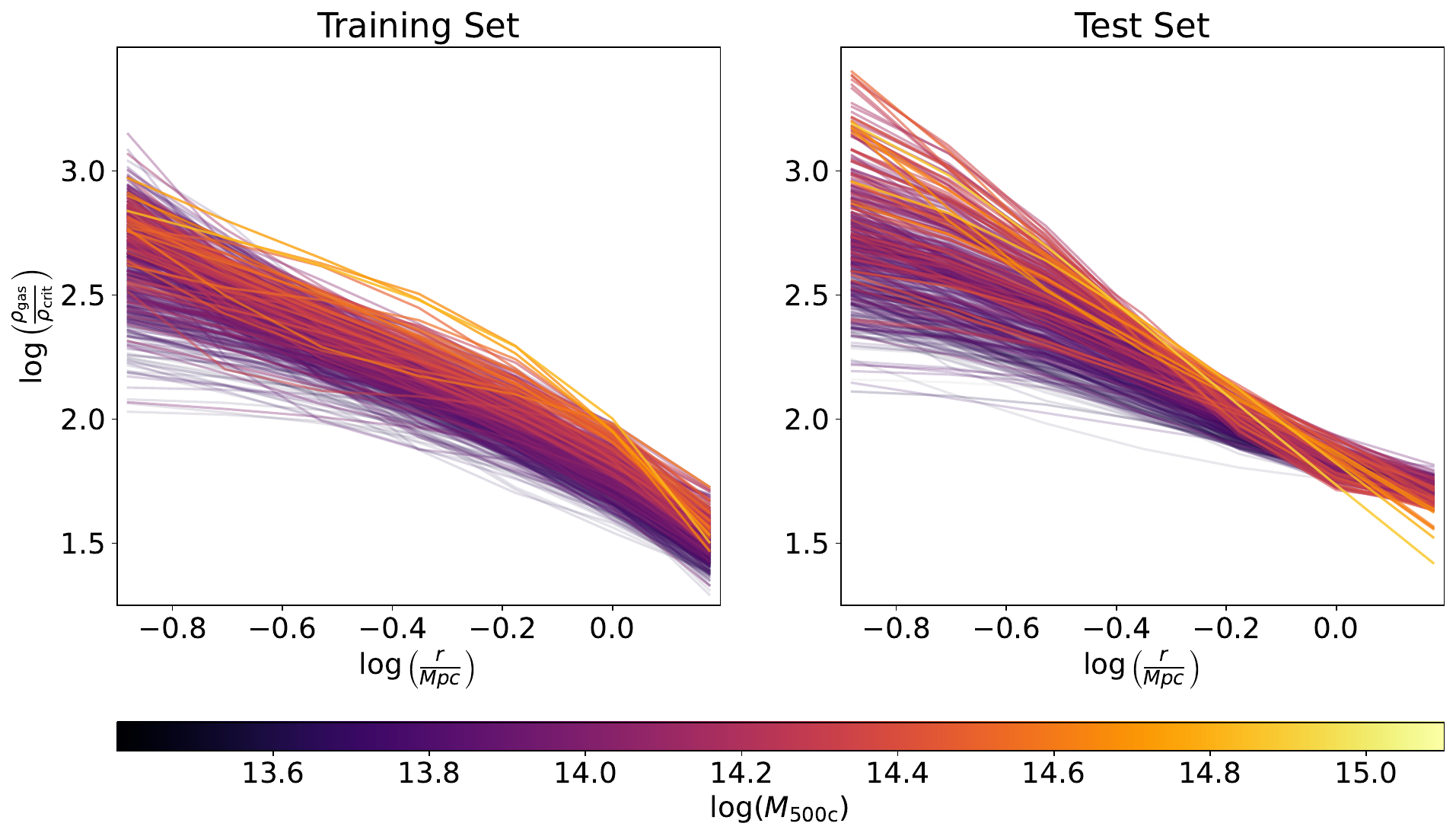} 

 \includegraphics[width=0.9 \textwidth]{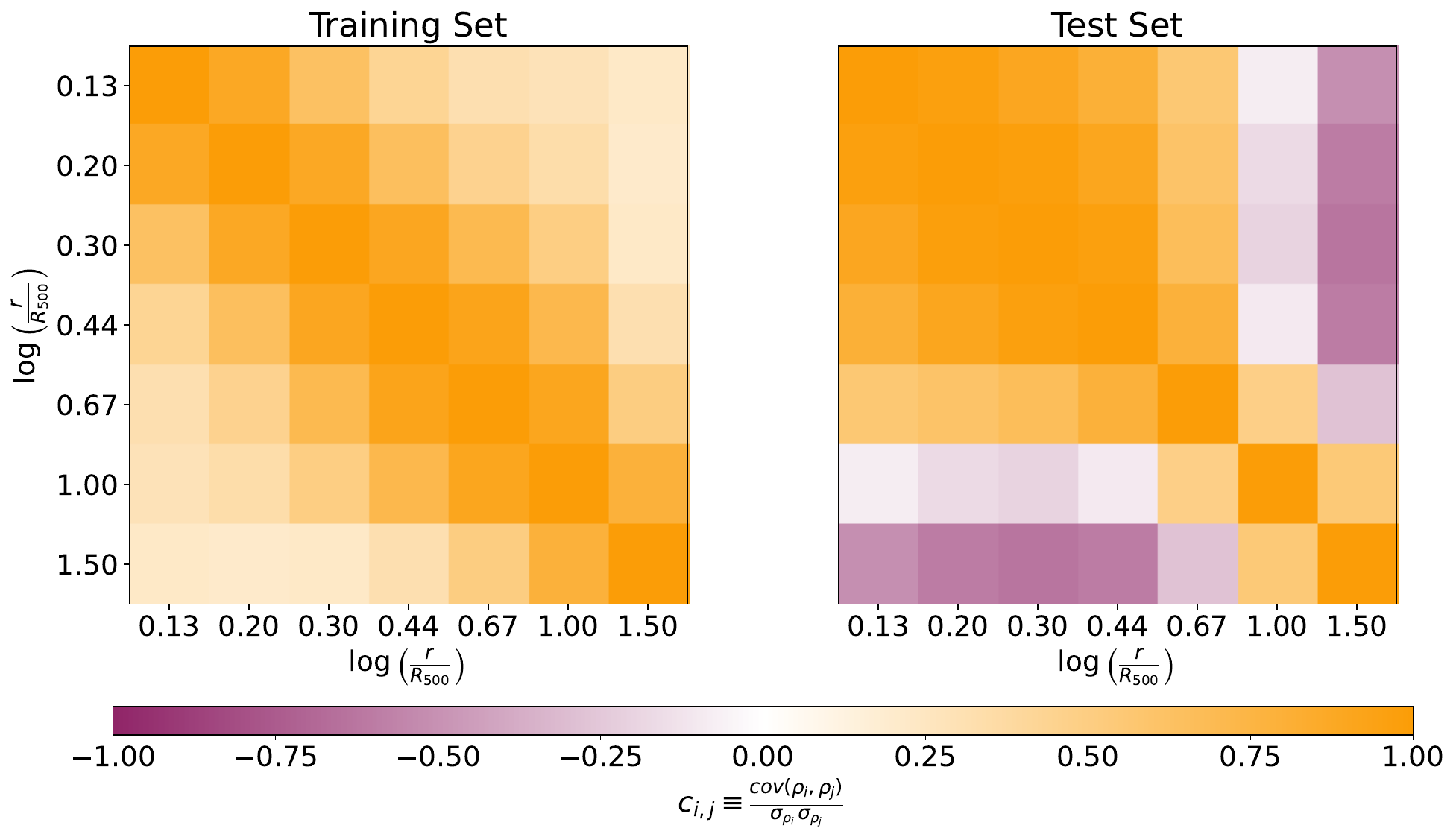} 
		\caption[]{Top: Representative sample of density profiles for the Training (left) and Test (right) Sets.  Profiles are colored by cluster mass.  The \texttt{Magneticum} profiles that form the Training Set are shallower in the core and less self-similar in the outskirts, an effect that is likely due to feedback differences between \texttt{Magneticum} and \tng. \\
        Bottom:  Correlation matrices showing how these same density profiles are correlated across logarithmically spaced radial bins.  The \texttt{Magneticum} profiles that comprise the Training Set have stronger positive correlations across disparate radial bins.  The \texttt{IllustrisTNG} clusters that comprise the Test Set, however, have very strong correlations among bins inside of $r\sim0.67R_{500}$, while the outskirts of these clusters are anticorrelated with the inner regions. }
       	\label{fig:sully}	
\end{figure*}

The Training and Test Sets were selected not for their similarities, but for their differences.  Because deep ML algorithms are susceptible to overspecialization \textemdash{} the tendency to base results on simulation-specific details \textemdash{} having a Test Set that is uniquely derived and potentially out of domain can be a powerful test for the robustness of the results.  For this research, both the underlying simulation and the process for generating mock observations from those simulations are unique, to provide a way to assess how the model responds to domain shift.

\subsubsection{Density Profiles}
\label{sec:densityprofiles}

The cluster samples have density profiles that are subtly visually different. Figure \ref{fig:sully} shows the cluster sample density as a function of cluster radius.  The Training Set profiles are shallower near the cluster cores; we theorize that this may be due to differences between the feedback models used in \magneticum{} and \tng{}, creating differences in the way gas is pushed from the cluster cores.  Feedback differences may also be the reason why the outskirts (near $r\approx R_{500}$) of the \magneticum{} clusters are less self-similar than the \tng{} cluster sample.  

In \magneticum{}, the lower-mass groups and clusters also have lower density normalization, and this is also likely due to the stronger feedback, which is more efficient in lower-mass halos, and has the effect of suppressing the gas mass.

Figure \ref{fig:sully} also shows the correlation matrices of these density profiles; the outskirts of the \magneticum clusters are more strongly correlated with the cluster cores, while the \tng clusters show an anti-correlation between outskirts and cores.

\subsubsection{$M-\mgas$ and $M-T$ Power Laws}
\label{sec:mgasT}
Assuming self-similarity, galaxy clusters' scaling relations should follow a simple power-law relationship given by:

\begin{equation}
    \log\left(\frac{M_{500}}{h^{-1}\,M_\odot}\right) = \alpha \log\left(P\right)+\mathcal{A}
\label{eq:PL}
\end{equation}
where $M_{500}$ is the cluster mass with an overdensity of $500$ times the critical density of the Universe, $P$ is the cluster observable such as gas mass (\mgas) or temperature ($T$).  The slope $\alpha$ and the normalization $\mathcal{A}$ are parameters that can be determined empirically.  The best fit parameters for the $\mgas-M$ and $T-M$ scaling relations are given in Table \ref{table:obs_compare}.  

\begin{figure}[]
	\centering
	\includegraphics[width=0.5 \textwidth]{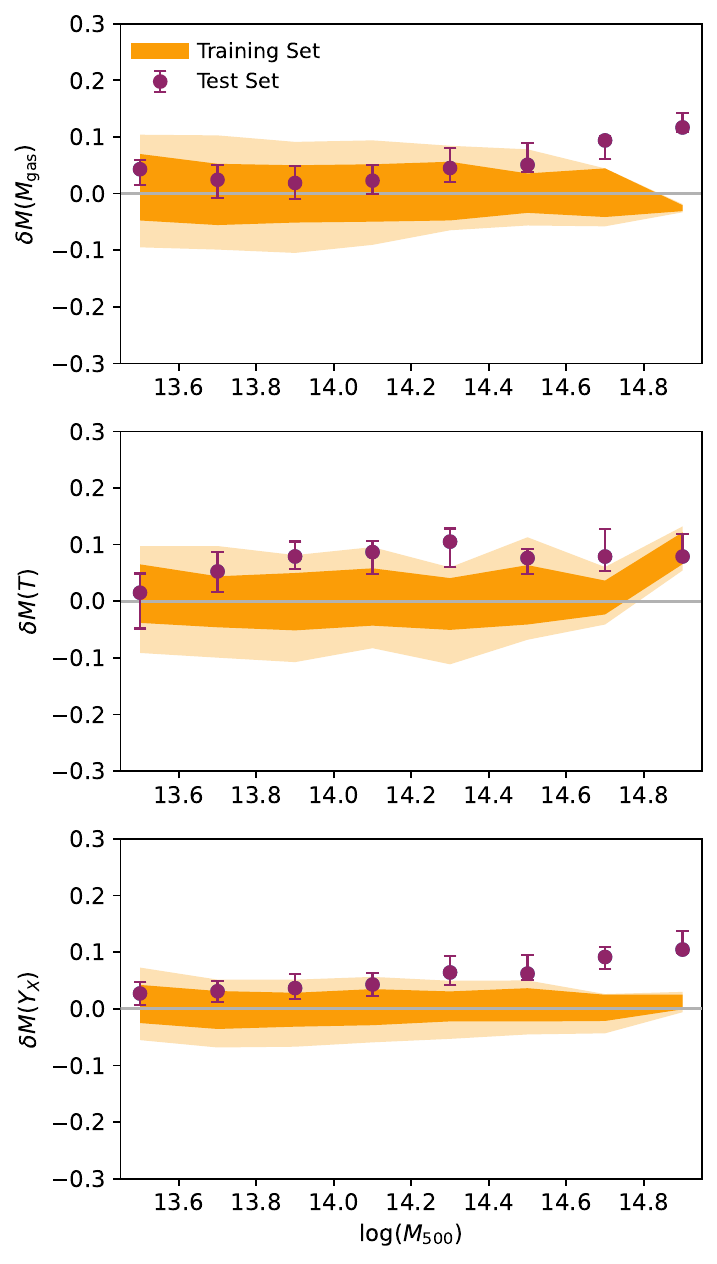} 
		\caption[]{Scatter associated with power law mass predictions for the $T-M_{500}$ (top), $M_\mathrm{gas}-M_{500}$ (middle), and $Y_X-M_{500}$ (bottom) scaling relations.  The Training Set (orange with 1- and 2-$\sigma$ bands) and Test Set (purple error bars with 1-$\sigma$ error bars) do not follow the same scaling relations, nor do they have the same intrinsic scatter.  The best-fit power law parameters for each simulation are compiled in Table \ref{table:obs_compare}.  When Training Set scaling parameters are adopted, scaling relation predictions on the Test Set have a mass-dependent bias. In practice, this is accounted for by employing an unbiased proxy (such as weak lensing mass estimates) to calibrate the underlying bias.}
       	\label{fig:powerlaw}	
\end{figure}

Figure \ref{fig:powerlaw} illustrates the differences between the \magneticum{} and \tng power laws; adopting the \magneticum{} power law best-fit parameters and applying them to the \tng{} cluster observations without calibration is obviously problematic, and it results in a mass-dependent bias.  This figure is for illustrative purposes only; for the remainder of this analysis, the simulations' power laws are fit independently.  For example, looking ahead to Figures \ref{fig:trainpred} and \ref{fig:testpred}, the mass errors as a function of \mgas and $T$ are compared against other mass proxies using independent power law fits for each simulation.

\begin{figure*}[]
	\centering
	\includegraphics[width=\textwidth]{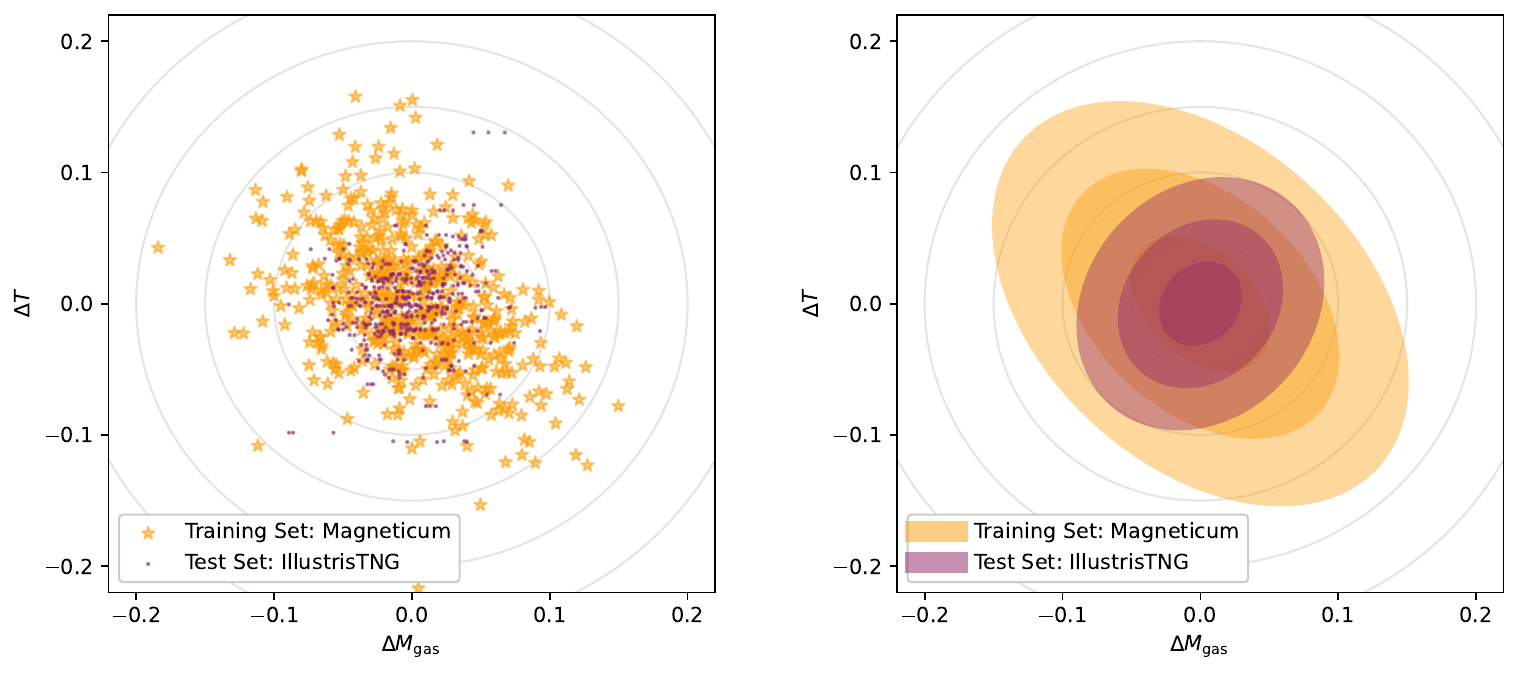} 
		\caption[]{The errors in gas-mass-based and temperature-based mass estimates of clusters in the Training and Test Sets. The left and right figures show the same information presented two different ways:  on the left, scatter points, and on the right, ellipses capturing the inner $1-$ and $2-\sigma$ of the same data.  In the Training Set (orange), the errors on mass estimates derived from gas mass ($\Delta\,M_\mathrm{gas}$) and those based on temperature ($\Delta\,T$) have anticorrelated errors, similar to the trend shown in Figure 4 of \cite{2006ApJ...650..128K}.  This anticorrelation is the primary underlying trend that is exploited in $Y_X$ to produce a lower-scatter mass proxy. In the Test Set (purple), however, these errors are positively correlated, resulting in a smaller net reduction in mass scatter by using $Y_X$.}
       	\label{fig:yx2}	
\end{figure*}

Though \cite{2022arXiv220511528P} finds that a broken power law scaling relation better describes the \tng{} cluster sample, we find that fitting the Test Set with a more flexible broken power law does not result in significant changes to the scatter or bias; this is likely due to the fact that the pivot point found by \cite{2022arXiv220511528P} is in the $13.7-14.0$ range for $\log(M_{500})$, which is near the lower end of the cluster sample used for this analysis.

\subsubsection{$M-Y_X$  Power Law}
\label{sec:yx}

\cite{2006ApJ...650..128K} found that in a sample of simulated galaxy clusters, errors on \mgas and $T$-based cluster mass estimates are anticorrelated.  This led to a robust, low-scatter composite mass proxy, $Y_X$, which takes advantage of a fundamental plane in 3-dimensional $M-\mgas{}-T$ space.  $Y_X$ is defined as
\begin{equation}
    Y_X \equiv \mgas \times T,
    \label{eq:yx}
\end{equation}
and using this mass proxy in lieu of \mgas{} or $T$ alone results in a $\sim38\%$ reduction in cluster mass error. Figure \ref{fig:yx2} shows a recreation of this finding.  While the \magneticum{} clusters do exhibit the expected negative correlation \citep[c.f. Figure 4 in][]{2006ApJ...650..128K}, the \tng clusters do not.  

Applying the broken power law as suggested by \cite{2022arXiv220511528P} does not rectify this.  As discussed in Section \ref{sec:mgasT}, this is likely because the pivot mass at which the two power laws transition is near the bottom of the \tng cluster sample used in this analysis. An explanation for why the \yx model reduces mass errors even though the expected error anticorrelation is not present is discussed in the section on the value of negative results (\ref{sec:negativeresults}).

\section{Methods: Deep Neural Network}
\label{sec:MLmethods}

\begin{deluxetable*}{l r l r r l}
\tablecaption{Standard and \multimodel Neural Network Architecture}
\label{table:architecture_NN}
\tablehead{
\colhead{Layer} &\colhead{Filters} &\colhead{Shape} & \colhead{Activation}  & \colhead{Dropout} & \colhead{Comments}
}
\startdata
input					&\nodata			& $N_\mathrm{cluster} \times 8$		& \nodata		& \nodata		& cluster global $T$ and $M_\mathrm{gas}$ profile\\
fully connected network  						&  2048 			& $N_\mathrm{cluster} \times 2048$	& leaky ReLU	& 30\%		& \nodata\\
fully connected network 						&  1024 			& $N_\mathrm{cluster} \times 1024$	& leaky ReLU	& 30\%		& \nodata\\
fully connected network 						&  512 			& $N_\mathrm{cluster} \times 512$	& leaky ReLU	& 30\%		& \nodata\\
fully connected network 						&  256 			& $N_\mathrm{cluster} \times 256$	& linear	& 0\%		& \nodata\\
output &  1 				& $N_\mathrm{cluster}  \times 1$		& \nodata	& 0\%		& predicted cluster mass\\
\enddata
\tablecomments{Figure \ref{fig:NN} shows a schematic version of this architecture.}

\end{deluxetable*}


\begin{deluxetable*}{l r l r r l}
\tablecaption{Deep Reconstruction-Regression Network}
\label{table:architecture_drrn}
\tablehead{
\colhead{Layer} &\colhead{Filters} &\colhead{Shape} & \colhead{Activation}  & \colhead{Dropout} & \colhead{Comments}
}
\startdata
\textbf{Encoder Architecture:} & & & & & \\
encoder input					&\nodata			& $N_\mathrm{cluster} \times 8$		& \nodata		& \nodata		& cluster global $T$ and $M_\mathrm{gas}$ profile\\
fully connected network  						&  2048 			& $N_\mathrm{cluster} \times 2048$	& leaky ReLU	& 30\%		& \nodata\\
fully connected network 						&  1024 			& $N_\mathrm{cluster} \times 1024$	& leaky ReLU	& 30\%		& \nodata\\
fully connected network 						&  512 			& $N_\mathrm{cluster} \times 512$	& leaky ReLU	& 30\%		& \nodata\\
fully connected network 						&  256 			& $N_\mathrm{cluster} \times 256$	& leaky ReLU	& 0\%		& \nodata \\
encoder output$\dagger$ &  3 				& $N_\mathrm{cluster}  \times 3$		& linear	& 0\%		& layer output = latent space\\
\tableline
\textbf{Decoder Architecture:} & & & & & \\
decoder input$\dagger$ &  3 				& $N_\mathrm{cluster}  \times 3$		& linear	& 0\%		& same layer as encoder output\\
fully connected network 						&  256 			& $N_\mathrm{cluster} \times 256$	& leaky ReLU	& 30\%		& \nodata\\
fully connected network 						&  512 			& $N_\mathrm{cluster} \times 512$	& leaky ReLU	& 30\%		& \nodata\\
fully connected network 						&  1024 			& $N_\mathrm{cluster} \times 1024$	& leaky ReLU	& 30\%		& \nodata\\
fully connected network  						&  2048 			& $N_\mathrm{cluster} \times 2048$	& linear	& 0\%		& \nodata\\
decoder output					&\nodata			& $N_\mathrm{cluster} \times 8$		& \nodata		& \nodata		& cluster global $T$ and $M_\mathrm{gas}$ profile\\
\tableline\textbf{Regressor Architecture:} & & & & & \\
regressor input$\dagger$ &  3 				& $N_\mathrm{cluster}  \times 3$		& linear	& 0\%		& same layer as encoder output\\
fully connected network  						&  2048 			& $N_\mathrm{cluster} \times 2048$	& leaky ReLU	& 30\%		& \nodata\\
fully connected network 						&  1024 			& $N_\mathrm{cluster} \times 1024$	& leaky ReLU	& 30\%		& \nodata\\
fully connected network 						&  512 			& $N_\mathrm{cluster} \times 512$	& leaky ReLU	& 30\%		& \nodata\\
fully connected network 						&  256 			& $N_\mathrm{cluster} \times 256$	& leaky ReLU	& 0\%		& \nodata \\
regressor output &  1 				& $N_\mathrm{cluster}  \times 1$		& linear	& 0\%		& predicted cluster mass\\
\enddata
\tablecomments{The three layers marked with $\dagger$ are the same; the encoder output, decoder input, and regressor input are a shared, 3-dimensional latent space.  Figure \ref{fig:drrn} shows a schematic version of this architecture.}
\end{deluxetable*}


Deep learning is a class of ML techniques designed to take advantage of subtle patterns and correlations in data.  The simplest deep learning models are fully connected neural networks, which process input through multiple unseen (or ``latent'') layers and can be trained to approximate complex functions. 

As discussed in Section \ref{sec:mgasT}, cluster global properties such as \mgas and $T$ straightforwardly map to cluster mass through a simple power law.  However, the details of the cluster mass profile contain additional, potentially valuable information about the assembly history, dynamical state, and environment of the cluster \citep[see][and references therein]{2025ApJ...985..212S}.  These factors contribute to the intrinsic scatter in the $\mgas-M$ and $T-M$ scaling relations and also leave imprints in the cluster gas mass profiles, and these subtle signals could potentially be used by deep learning to improve model accuracy and precision.

The aim of this research is not just to find and harness these signals, but also to understand model robustness, to explore the impact of domain shift, and to quantify the efficacy of domain adaptation methods under a realistic domain shift.  We explore three deep learning approaches that estimates cluster mass from mock cluster observations, described below and summarized in Table \ref{table:results}.  

The models that are described in the sections below increase in complexity, with the Neural Network (NN, \S\ref{sec:NN}) having no domain adaptation built in, the \multimodel Neural Network (SANN, \S\ref{sec:NN}) having domain adaptation built into the Training Data, and the Deep Reconstruction-Regression Network (DRRN, \S\ref{sec:drcn}) using a model designed to manage domain adaptation.

\begin{table*}[t] 
\caption{Summary of experiments, in increasing in complexity}
\label{table:results}
\begin{center}
\begin{tabular}{| R{4cm} |  R{4.5cm} | R{1.5cm} | R{2.5cm} | R{2.5cm} |}
\hline
\hline
Name & Method & Color & Training/Val Results & Test Results \\
\hline
\hline
	$M-M_\mathrm{gas}$ Power Law (\S\ref{sec:mgasT})
	& Power Law, calibrated
	& Lime
	& $\sigma=11.94\%$ \vfill $b:=0.00$   &  $\sigma=6.83\%$ \vfill $b:=0.00$  \\
\hline
	$M-T$ Power Law\vfill(\S\ref{sec:mgasT})
	& Power Law, calibrated
	& Yellow
	& $\sigma=11.93\%$ \vfill $b:=0.00$   &  $\sigma=9.94\%$ \vfill $b:=0.00$  \\
\hline
	$M-Y_\mathrm{X}$ Power Law\vfill (\S\ref{sec:yx})
	& Power Law, calibrated
	& Teal
	& $\sigma=7.38\%$ \vfill $b:=0.00$   &  $\sigma=5.39\%$ \vfill $b:=0.00$  \\
\hline
	Standard NN\vfill (\S\ref{sec:NN})
	& Deep Learning, uncalibrated
	& Blue
	& $\sigma=6.10\%$ \vfill $b=0.00$   &  $\sigma=7.45\%$ \vfill $b=-0.02$  \\
\hline
	  \multimodel NN\vfill (\S\ref{sec:trainingset} \& \ref{sec:NN})
	& Deep Learning, marginalized over many models
	& Purple
	& $\sigma=7.83\%$ \vfill $b=-0.01$   &  $\sigma=8.39\%$ \vfill $b=-0.02$  \\
\hline
	  DRRN\vfill (\S\ref{sec:drcn})
	& Deep Learning with domain adaptation
	& Black
	& $\sigma=7.67\%$ \vfill $b=-0.01$   &  $\sigma=7.59\%$ \vfill $b=-0.03$  \\
\hline
\hline
\end{tabular}
\end{center}
\end{table*}

\subsection{Neural Network Inputs}
\label{sec:NNinputs}

\begin{figure}[]
	\centering
	\includegraphics[width=0.5\textwidth]{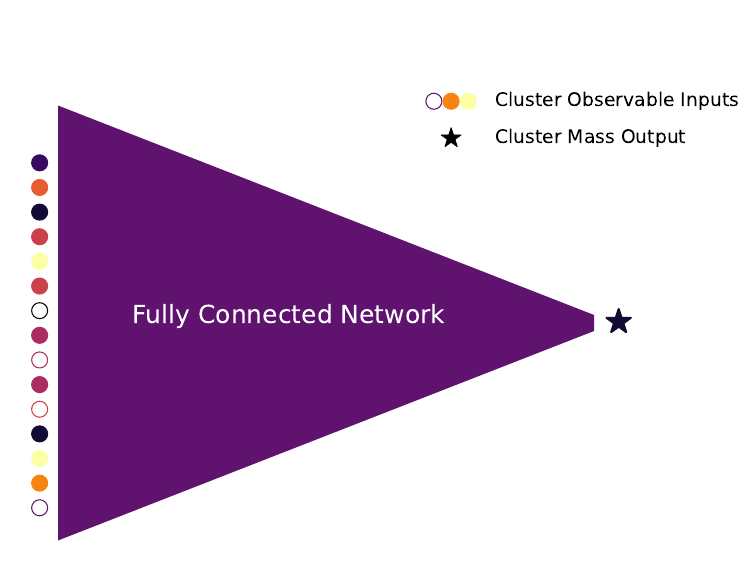} 
		\caption[]{The standard NN follows a very basic, fully connected architecture with leaky ReLU activation functions and moderate dropout after early layers.  The model learns from cluster gas mass (\mgas), cluster gas mass profiles ($M_\mathrm{g}(r)$), and cluster mass-weighted temperature ($T$) to output a cluster mass estimate.  The full architecture of the model is given in Table \ref{table:architecture_NN}.}
       	\label{fig:NN}	
\end{figure}

Each network has a 9-parameter input that includes the cluster global mass-weighted temperature, $\log(T)$, total observed gas mass $\log(M_\mathrm{gas})$, and each of the 7 radially binned gas masses described in Section \ref{sec:densityprofiles} and shown in Figure \ref{fig:sully}.  Deep Learning models tend to train faster and give better results when the input and output values are bound roughly between 0 and 1, so all inputs are linearly rescaled so that the Training Set values are exactly between 0 and 1.  The \multimodel Training, Validation, and Test Sets are rescaled with identical parameters, resulting in input values that may extend slightly beyond [0,1].  The models are trained to output a single parameter, the cluster mass $\log(M)$, unless noted otherwise.  The cluster mass is linearly scaled in the same way as the inputs: the Training Set is exactly between 0 and 1, and the other sets use the same linear scaling parameters.

\subsection{Standard and \multimodel Neural Networks}
\label{sec:NN}

For this first and simplest deep learning experiment, we utilize a straightforward fully connected neural network with a leaky rectified linear unit \citep[leaky ReLU, e.g.,][]{2015arXiv150500853X} activation function and moderate dropout in early layers, described in detail in Table \ref{table:architecture_NN}.

The model is built using \texttt{Keras} \citep{chollet2015} with a \texttt{TensorFlow} \citep{45381} backend.  It is trained for 300 epochs using the stochastic gradient descent optimizer \citep{bottou2010large} and a mean absolute error (MAE) loss function. The mean squared error (MSE) of the Validation Set is calculated in real-time, and the epoch with the lowest MSE is selected as the final model.   

While the Standard Neural Network is trained with the Standard Training Set, the \multimodel Neural Network (SANN) is trained using the \multimodel Training Set.  All other training and model selection details are identical.  Figure \ref{fig:NN} shows a schematic diagram of the NN and SANN architecture, and full details are given in Table \ref{table:architecture_NN}.

\subsection{Deep Reconstruction-Regression Network}
\label{sec:drcn}

A Deep Reconstruction-Classification Network (DRCN), first proposed by \cite{2016arXiv160703516G}, is a semi-supervised deep model engineered to address challenges of domain shift.  It is a similar approach to the Domain Adversarial Neural Network (DANN) proposed in \cite{ganin2016domain}, which uses a classifier as a tension against the training and testing sets being treated differently.  The DRCN, instead, uses an autoencoder to supply that pressure. Both DNNs and DRCNs use auxiliary tasks to enable improved alignment of the training and testing data in the latent space of the model. Similarly, distance-based domain adaptation methods, such as Maximum Mean Discrepancy \citep[MMD, e.g.][]{gretton2012kernel, 2021MNRAS.506..677C}, map the training and Test Data onto a similar latent space by minimizing a distance metric between these distributions during training.

\begin{figure*}[]
	\centering
	\includegraphics[width=0.72\textwidth]{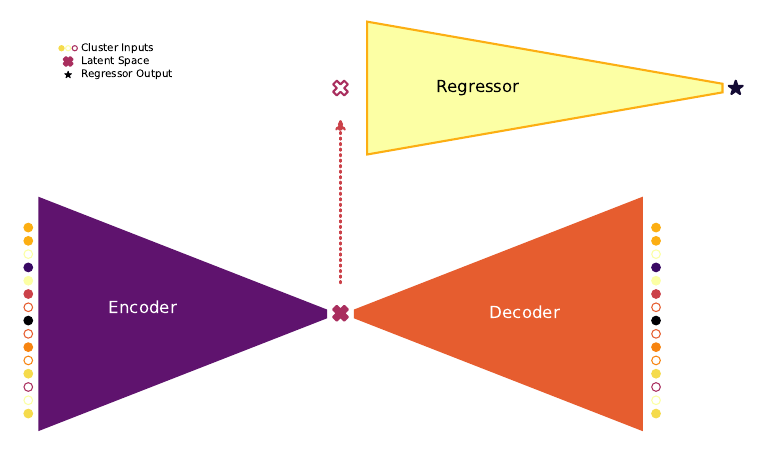} 
		\caption[]{A Deep Reconstruction-Classification Network, first proposed in \cite{2016arXiv160703516G}, includes two networks that share a latent space.  The first network is shown in the bottom half of the figure and includes an unsupervised autoencoder, shown by the encoder network (purple), the latent parameters (magenta X), and the decoder network (orange).  Because this portion of the network is unsupervised and simply learns to summarize and re-create the input signal, it does not need labeled data; this portion of the network is trained using both the labeled Training Data and the unlabeled Test Data.  The second network shares the encoder and latent value(s), but then moves into a regression network \citep[or, more traditionally, a classification network as in][]{2016arXiv160703516G}, shown in yellow, which is trained with labeled data only to output the estimate (purple star).  Training consists first of training the first network (encoder plus decoder) for one epoch, then the second network (encoder plus regressor) for one epoch.  Training cycles through training one network and then the other.  By alternating between training these two networks and their shared encoder, the model learns not only how to label the data, but also how to describe the Test Data, which may be out of domain.  Because of this two-pronged approach to training, \cite{2016arXiv160703516G} argues that the model is more robust to domain shift.\\ }
       	\label{fig:drrn}	
\end{figure*}

Our network is modeled on \cite{2016arXiv160703516G}'s network but uses a regressor rather than a classifier, and is summarized in Table \ref{table:architecture_drrn}, shown in Figure \ref{fig:drrn}.  The model has an unsupervised autoencoder \citep[AE,][]{hinton2006reducing} network paired with a supervised regression network; these two networks share an encoder and latent space.  For each epoch of training, the AE portion is first trained to reconstruct the Training and Test Set signals, then the Regression Network is trained to estimate the Training Set cluster masses.  The training proceeds, updating first the AE, then the Regression Network. 

The model is built using \texttt{Keras} \citep{chollet2015} with a \texttt{TensorFlow} \citep{45381} backend.  It is trained for 700 epochs using the stochastic gradient descent optimizer \citep{bottou2010large} and a mean absolute error (MAE) loss function for both the AE and Regressor networks. It is trained with the Standard Training Set (input data and labels) and the Test Set (input data only) as described above.  The MSE of AE signal reconstruction is calculated on the fly, and the model with the lowest autoencoder MSE is selected.

\begin{figure*}[]
	\centering
	\includegraphics[width=0.9\textwidth]{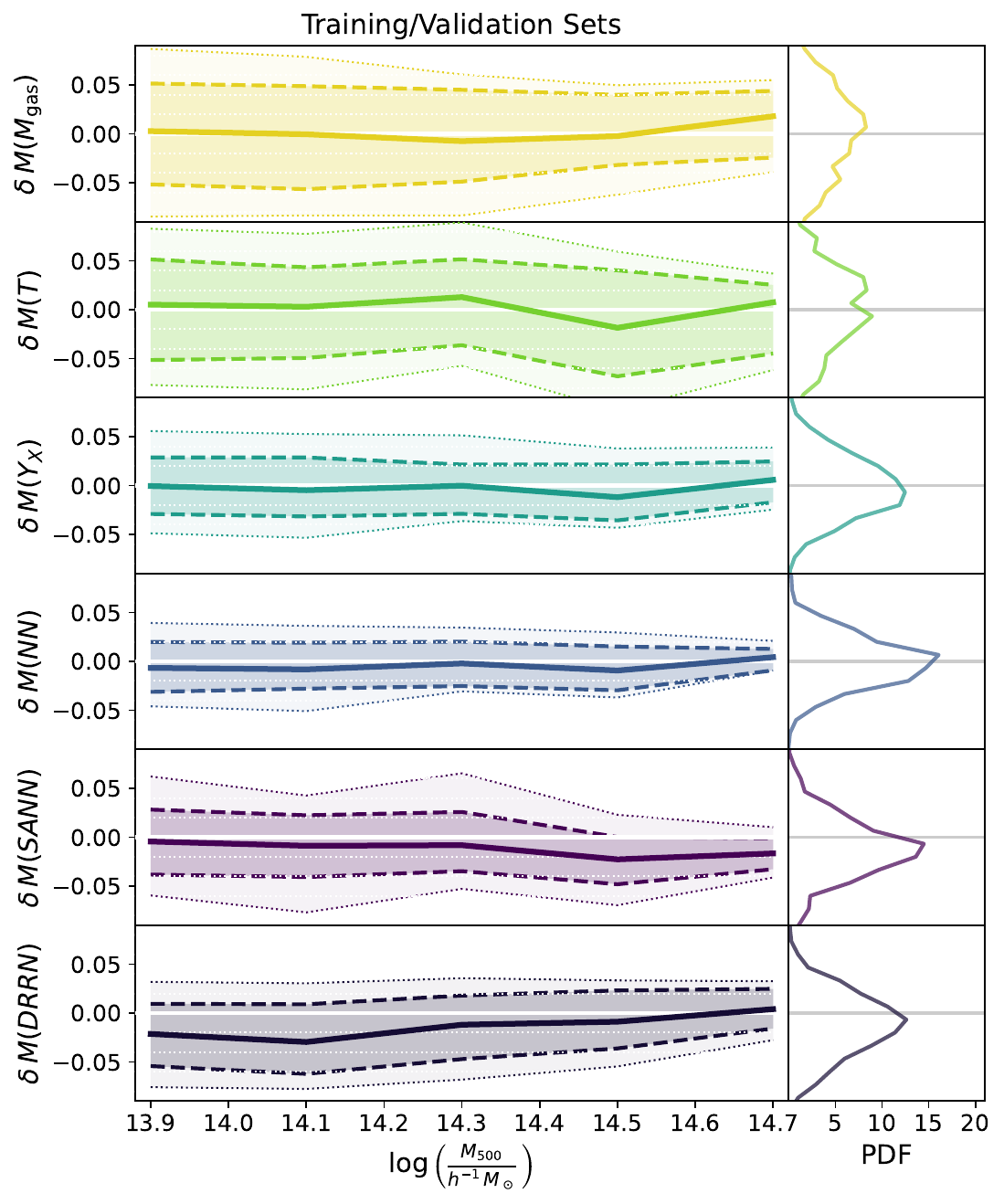} 
		\caption[]{Left column: Mass-binned distribution of errors $\delta$ [median (solid curve), 68\% (dashed curve), and 95\% (dotted curve)] for the Validation Set for the three traditional statistical methods [top to middle: $\Mgas$ scaling relation (yellow), $T$ scaling relation (lime green), and \yx scaling relation (teal)] and the three deep learning methods [middle to bottom: Standard Neural Network (NN, blue), \multimodel Neural Network (SANN, purple), and the Deep Reconstruction-Regression Network (DRRN, black)].  White solid lines trace $\delta=0.00$ and white dotted lines trace $\delta=\pm0.02, \pm0.04, \pm0.06$, for visual reference.  Right column:  PDF of errors across all mass ranges.  From this result, it appears that the NN significantly outperforms all traditional statistical methods, but we will see in Figure \ref{fig:testpred} that this does not hold true for the out-of-domain Test Set.}
       	\label{fig:trainpred}	
\end{figure*}

\begin{figure*}[]
	\centering
	\includegraphics[width=0.9\textwidth]{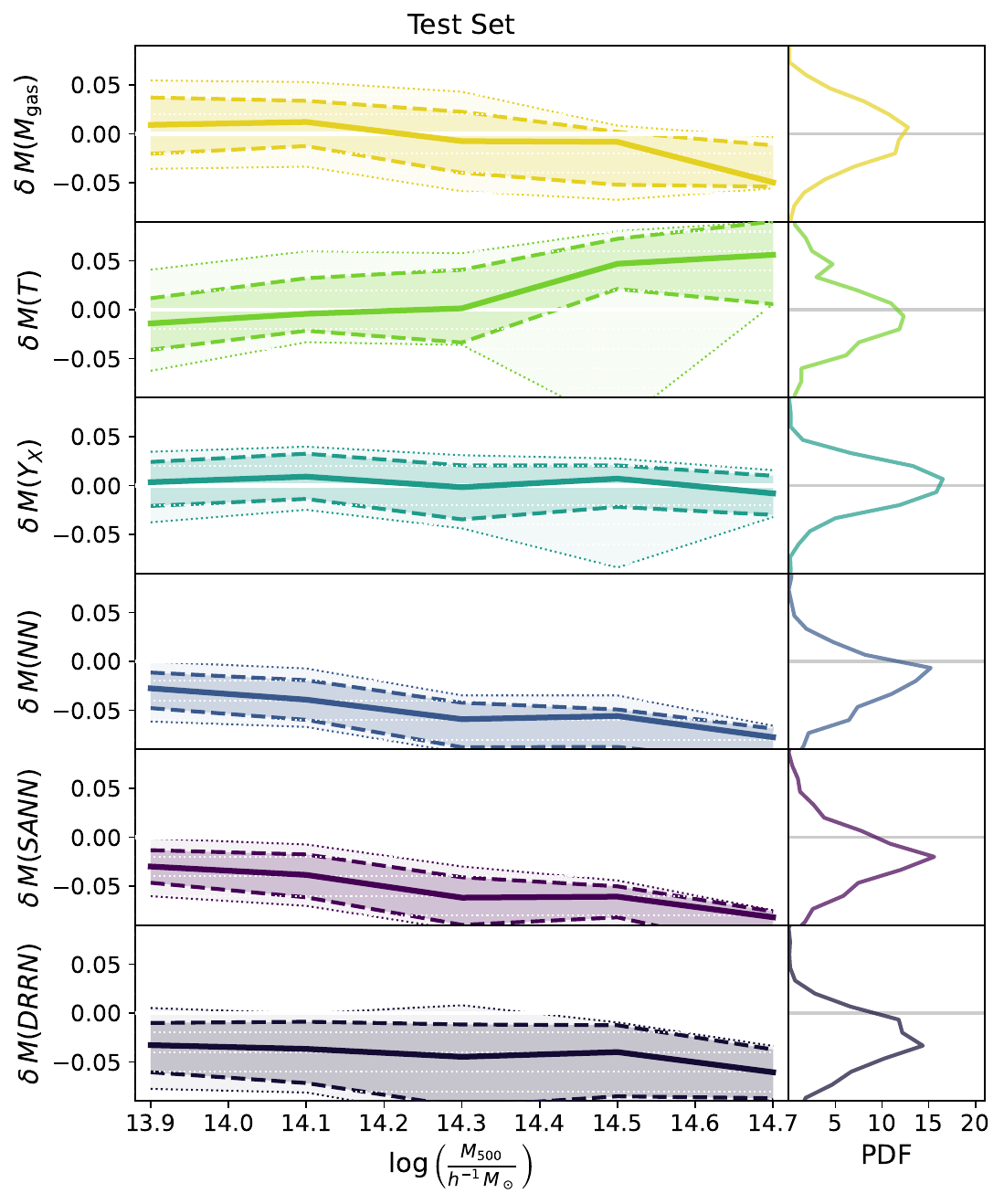} 
		\caption[]{Same as Figure \ref{fig:trainpred}, but for the Test Data.  While the top three traditional statistical methods inherently include calibration, the bottom three deep learning methods do not, resulting in a significant underestimate of cluster masses across the entire mass range.  This is especially pronounced in the PDF of errors in the right column, which are all centered well below the gray 0 line.}
       	\label{fig:testpred}	
\end{figure*}

\begin{figure*}[]
	\centering
	\includegraphics[width=\textwidth]{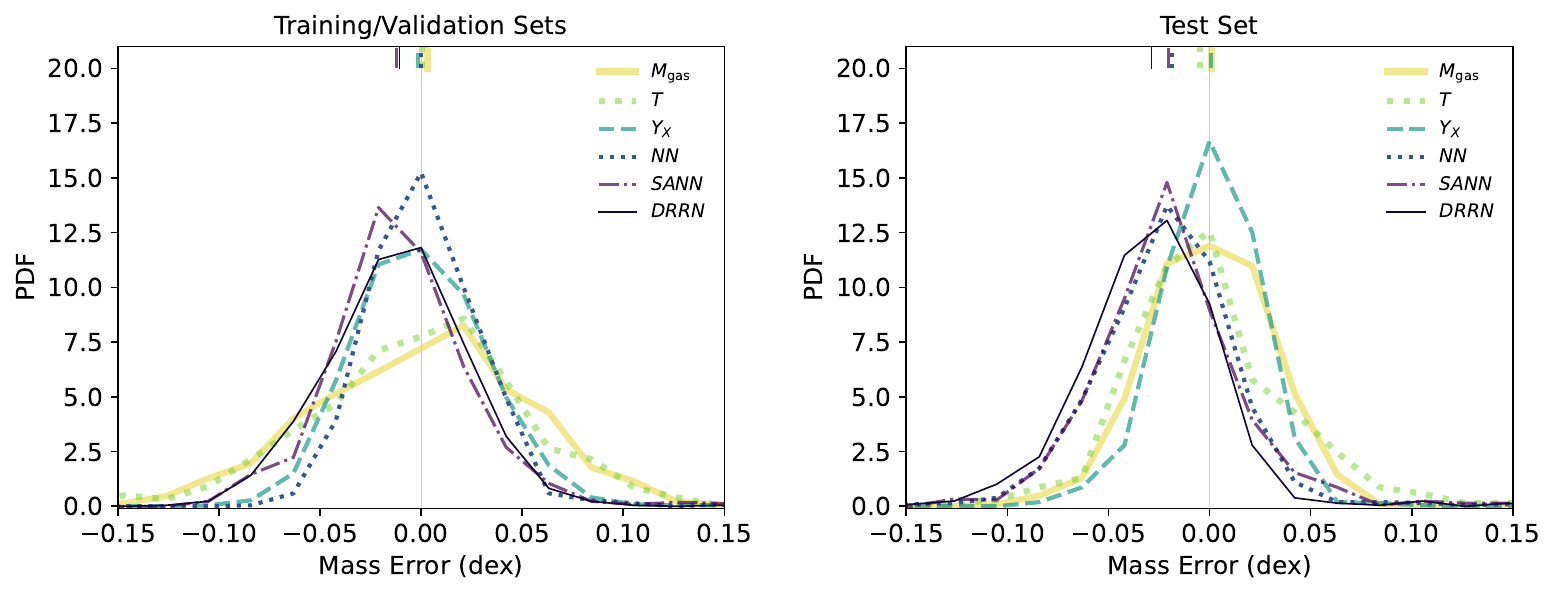} 
		\caption[]{PDF of errors (curves) and median error (small dashes at the top of the figure frame) for the three traditional statistical methods [$\Mgas$ scaling relation (yellow solid), $T$ scaling relation (lime green dotted), and \yx scaling relation (teal dashed)] and the three deep learning methods [standard neural network (NN, blue dotted), neural network train with scatter augmentation (SANN, purple dash dotted), and the Deep Reconstruction-Regression Network (DRRN, black solid)].\\
        Left:  As expected, \yx outperforms $\mgas$ and $T$ scaling relations.  The NN outperforms \yx by $17\%$, resulting in a mass estimate error of $6.1\%$, as reported in Table \ref{table:results}. The SANN and DRRN models perform slightly worse than \yx, at 7.6\% and 7.7\% scatter, respectively.  While the error PDFs for the first four models are centered on zero, the SANN and DRRN models have median errors slightly below zero, as indicated by the dashes at the top of the figure.\\
        Right: The models' performance on the out-of-domain Test Set gives some indication of the robustness of each approach.  As before, \mgas and $T$ scaling relations underperform compared to \yx.  The three deep learning models are all biased low on the Test Data, and NN and SANN both underperform compared to \yx.  While the DRRN underperforms compared to \yx in both the Training and Test Sets, the results for the DRRN in both are very similar, indicating that it is more robust to domain shift, despite not being an improvement over the traditional statistical methods.\\}
       	\label{fig:errpdf}	
\end{figure*}

\section{Results \& Discussion}
\label{sec:results}

\subsection{Precision, Accuracy, and the Role of Calibration}

The bottom panel of Figure \ref{fig:powerlaw} highlights the role of calibration in scaling relations; without properly calibrating \yx{} on the Test Data, and instead relying on the calibration of the Training Data, the resulting mass estimates on the Test Set are biased high.  This is not a novel statistical concept, but it's one that is easy to forget in simulation-based inference: calibration is key, and it is dangerous to blindly trust that an ML model is properly calibrated for a data set it's never seen.

Figures \ref{fig:trainpred}, \ref{fig:testpred}, and \ref{fig:errpdf} again highlight the importance of calibration.  
Figure \ref{fig:trainpred} shows the mass-binned distribution of errors for the three scaling (Training and Validation Data) and three deep learning (Validation Data only) approaches.  Because the clusters shown in Figure \ref{fig:trainpred} are in domain compared to the data used for training or model fitting, the model has performed as expected and produces unbiased results.  Figure \ref{fig:testpred} tells a different story:  the scaling relations are calibrated, but the ML methods are not and produce biased results. Figure \ref{fig:errpdf}  is a PDF of errors for the validation and Test Sets, and it highlights this bias; on the Test Set, the deep learning models are biased.


We can draw three interesting conclusions by comparing the results on the Validation Set to those on the Test Set in Figure \ref{fig:errpdf}.  First, while the NN appears to be an improved method, showing a 17\% improvement in scatter on the Validation Set compared to \yx, the patterns that it has learned do not transfer, resulting in a significant, mass-dependent bias.  On the Test Set, the NN model is not only biased towards underprediction, but it also underperforms against \yx by 40\%.  

Second, the SANN model surprisingly does not outperform the NN model in terms of bias.  The \multimodel Training Set was engineered to encompass many physical models and scaling relations, with the goal of training a network that has marginalized over many possible mass-observable scalings.  But it did not work, indicating that there are differences that are beyond the scope of this simple approach to build flexibility into the model.  And, unfortunately, the Test Set remained out of domain.

Third, the DRRN, though it does not result in any remarkable improvement of error or bias, does have the benefit of ``truth in advertising'' and the results on the Validation and Test Sets are very similar.

We should conclude from this that there are use cases where deep learning regression can be extremely powerful, but that calibration, accuracy, and bias all require extra attention when testing on a data set that may be subtly out of domain.

\subsection{Model Interpretation}
\label{sec:interp}

In their guidance document ``Constructing Impactful Machine Learning Research for Astronomy: Best Practices for Researchers and Reviewers,'' \cite{2023arXiv231012528H} advocate for interpreting, diagnosing, and visualizing models and for exploring models' limits and scope.  We use saliency maps to explore the NN and SANN, and we apply a latent space visualization to explore the DRRN, to identify why they succeed on the Training Set, and to understand why they might fail on the Test Set.\\

\subsubsection{NN and SANN Saliency}

Saliency maps \citep{2013arXiv1312.6034S} are a technique for visualizing and understanding the sensitivity of a deep model's output on small changes to the input.  While they are traditionally applied to images, producing a heat map of pixel importance, the approach is straightforward to also apply to deep neural networks.  

The saliency $\mathcal{S}_i$ of the $i^{th}$ input is given by
\begin{equation}
    S_i \equiv \frac{\partial M}{\partial p_i},
\end{equation}
where $p_i$ denotes the $i^{th}$ input to the network and $M$ is the network output, the cluster mass.

\begin{figure*}[h!]
	\centering
	\includegraphics[width=\textwidth]{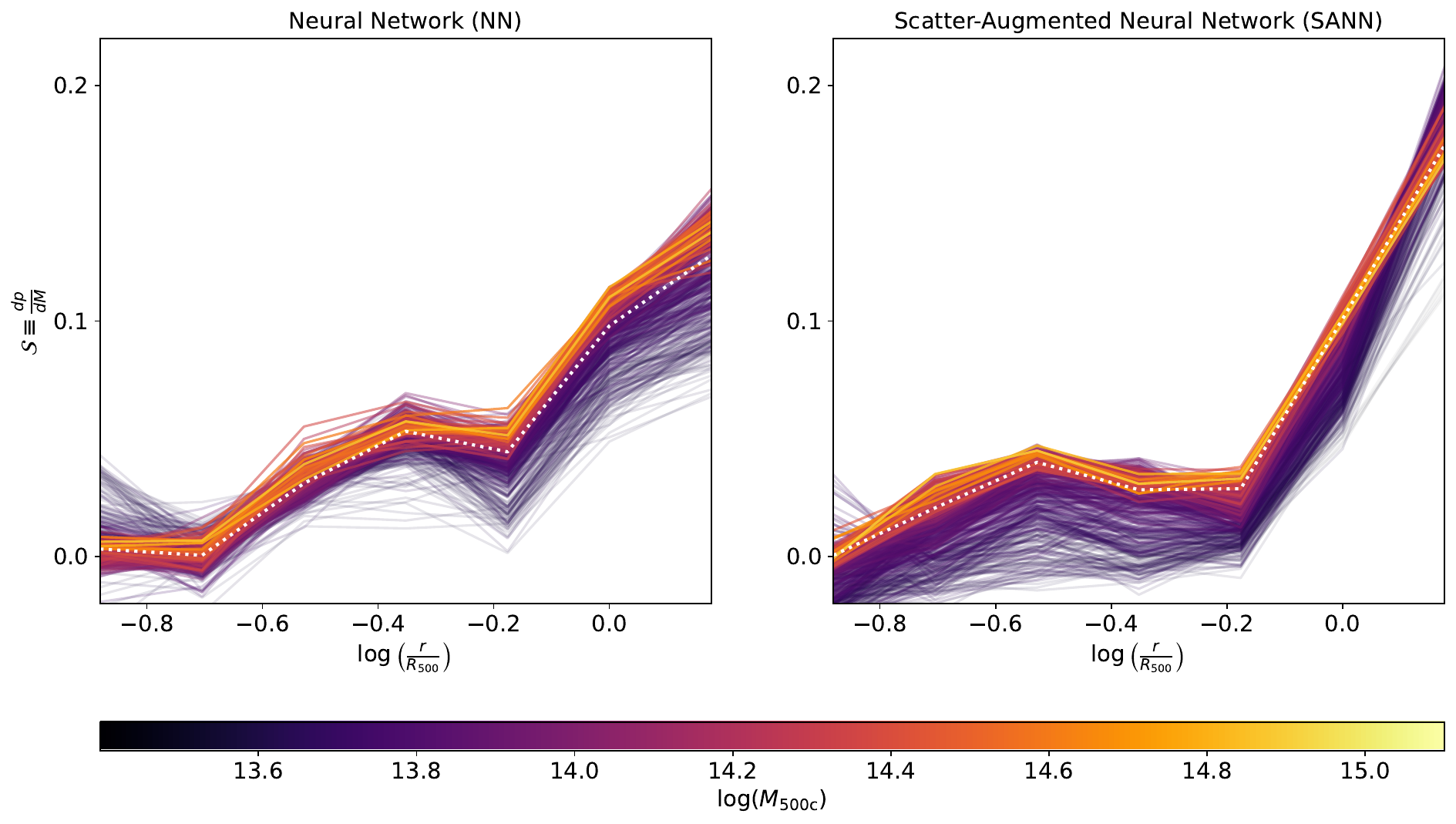} 
		\caption[]{Representative sample of saliency ($\mathcal{S}$) for the Test Set gas mass profiles as a function of distance from the cluster center in $R_\mathrm{500}$ units for the Neural Network (left) and \multimodel Neural Network (right). Profiles are colored by cluster mass. Both models are insensitive to cluster cores, which is unsurprising \textemdash{} the gas mass and luminosity of the inner regions of clusters are known to be only weakly correlated with cluster mass.  Both models pay more attention to cluster outskirts, with mass estimates highly sensitive to the gas mass at $R_\mathrm{500}$ and beyond.  Saliency trends are not strongly correlated with cluster mass; the average saliency (white dotted) captures the shape of the overall saliency trends across cluster masses.}
       	\label{fig:saliency}	
\end{figure*}

Figure \ref{fig:saliency} shows the saliency of the gas mass profile.  The model virtually ignores the inner regions of the cluster, with a mean saliency near zero.  This is an encouraging result that the model has learned something meaningful: the gas mass in cluster cores is only weakly correlated with mass.  \cite{Mantz:2010aa} found that excising the cores from X-ray analyses led to smaller errors in mass estimates, a result that was also found using deep learning interpretability  \citep[e.g.,][]{2019ApJ...876...82N, 2023MNRAS.524.3289H}.  Therefore, it's unsurprising to see a relatively small saliency in the inner region.  The model pays the most attention to the outer regions, at $R_\mathrm{500}$ ($\log(r/R_\mathrm{500})=0$) and beyond.  

This is a bit misleading, however, because each gas mass bin was normalized individually, linearly scaling to set the minimum value to 0 and the maximum to 1 (see Section \ref{sec:NNinputs}).  If we look back to Figure \ref{fig:sully}, we see that the gas density falls off sharply at large distances from the cluster center and has a very small scatter compared to the middle and inner cluster regions.  So while these regions are informative, it is actually the middle regions where gas mass is large, the saliency denominator $\delta p$ represents a significant physical change in the cluster, and saliency is substantial, which is the most important single feature to inform cluster mass.  It's important to note here that the NN and \multimodel NN are learning complex correlations among the inputs (if they did not, they would not be able to outperform \yx on the Validation Set, as shown in Figure \ref{fig:errpdf}).  The Saliency analysis shown here does not provide any information about those higher-dimensional correlations.\\

\subsubsection{DRRN Latent Space}

\begin{figure*}[]
	\centering
	\includegraphics[width=\textwidth]{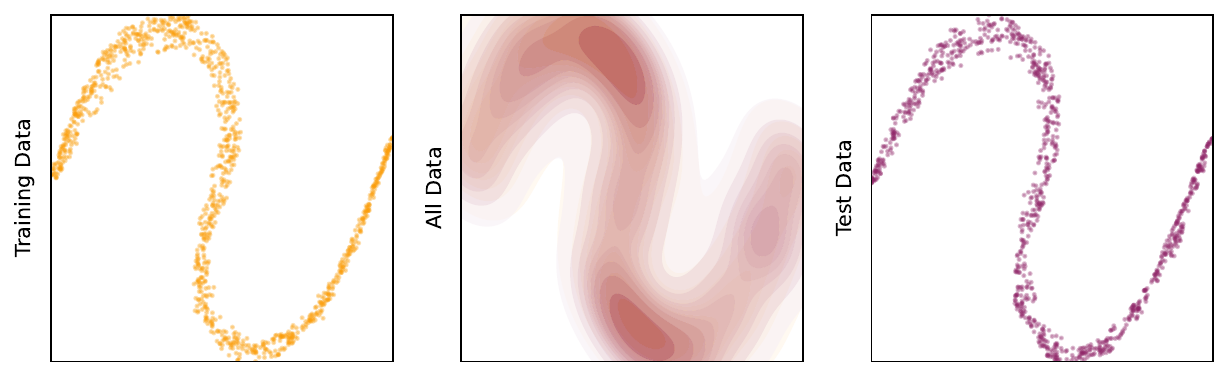} 
		\caption[]{The t-distributed stochastic neighbor embedding (t-SNE) is a statistical method for visualizing high-dimensional data; this figure shows the DRRN latent space projected into 2 dimensions.  Because this space has arbitrary units, all three panels are visualized with the same x- and y-extent, but without ticks or numerical labels.  The Training Data scatter points (left panel), and Test Data scatter points (right panel) inhabit nearly identical regions of this space, highlighted by the smoothed distributions of each data set (center panel).  From a visual inspection, we conclude that the DRRN encoder mapped the Training and Test Data onto a similar space, as desired, but unfortunately, this is not sufficient to outperform the \yx scaling relation, in terms of accuracy and precision, as shown in Figure \ref{fig:errpdf}.}
       	\label{fig:tsne}	
\end{figure*}

To assess whether the testing data are out of domain, we apply a t-distributed stochastic neighbor embedding \citep[t-SNE,][]{van2008visualizing} approach.  This statistical method can be used to visualize high-dimensional data in lower dimensions.  The t-SNE analysis of the latent values are shown in Figure \ref{fig:tsne}.  The exact output of this method is highly dependent on the random initialization, making the result scientifically, but not numerically, reproducible.  Therefore, details of Figure \ref{fig:tsne} should not be over-interpreted.  

We see that the model is successful in mapping the Training and Test Sets onto the same latent space.  Unfortunately, this is not sufficient for the trained model to be accurate, precise, and calibrated off the shelf.  We conclude, then, that testing whether data sets inhabit the same latent space is unlikely to be a sufficient proxy for model accuracy and precision on out-of-domain Test Sets.

\clearpage
\subsection{The Value of Negative Results}
\label{sec:negativeresults}

Machine Learning results are traditionally evaluated on some improvement to a benchmark task \textemdash{} the higher the performance improvement, the more valuable the result.  But \cite{2024arXiv240603980K} argue that negative results can be a powerful tool for building a comprehensive understanding of the field and for identifying important and unsolved problems.  They highlight two examples of negative ML results \citep[][]{bengio1994learning, 2013arXiv1312.6199S} that ultimately led to innovative ML solutions \citep[][]{hochreiter1997long, goodfellow2014explaining}, opening new subfields of active research.  Few venues for sharing negative ML results exist; one is the \textit{I Can’t Believe It’s Not Better!} workshop, which focuses entirely on the sharing and discussion of negative results to encourage learning and to advance the field through transparency.  

For ML applications in Astronomy, \cite{2023arXiv231012528H} advocate that ML models should be explored and their limits understood, and also that lessons learned \textemdash{} the failed attempts and negative results \textemdash{} should be described and discussed.  

Making the connections between valuable signals and physical meaning is an active area of research.  Current efforts to address this issue in astronomy include training interpretable-by-design neural networks \citep[e.g.,][]{2025ApJ...980..183W} and exploring field-level causes of biased inference \citep[e.g.,][]{2025ApJ...989..207B}.  The \yx results shown in Figures \ref{fig:yx2} and \ref{fig:errpdf} serve as a simple reminder of this: that valuable signals are not necessarily physical ones. In Figure \ref{fig:yx2}, we see that the Training Data $\Delta M_\mathrm{gas}$ and $\Delta T$ are anticorrelated, as expected from \cite{2006ApJ...650..128K}.  The Test Data, however, are not.  If these two errors are not anticorrelated, why does the right panel of Figure \ref{fig:errpdf} show that \yx has a smaller scatter than $M_\mathrm{gas}$ and $T$ individually?  Here, simple statistics are at play.  $\log(Y_x)=\log(M_\mathrm{gas})+\log(T)$, meaning that the \yx mass estimate is simply a multiplicative factor times the mean of the \mgas and $T$-based mass estimates, and the mean of two uncorrelated estimates will be, on average, closer to the truth than the individual estimates.  This variance reduction through averaging, and not some deeper underlying physical trend in the data of Figure \ref{fig:yx2}, is the reason why \yx works as a low-scatter mass proxy for the Test Data.

This argument extends to the machine learning results, which are based on training profiles that do not represent the Test Data (Figures \ref{fig:sully} and \ref{fig:yx2}) and which would need to be debiased with weak lensing to be useful (Figure \ref{fig:errpdf}).  Though significant effort was made to construct a Training Set that would encompass the Test Data (Section \ref{sec:trainingset}) and a t-SNE analysis showed that the Test Set was in domain (Fig \ref{fig:tsne}), still the model failed to generalize.\\

\section{Conclusions}
\label{sec:conclusions}

We have performed an experiment to explore the efficacy of two domain adaptation approaches for a realistic model of domain shift: estimating galaxy cluster masses when the Test Data are out of domain compared to the Training Data.

We developed Training and Validation Sets of mock galaxy cluster X-ray observations from the \magneticum simulation, including cluster mass-weighted temperature $T$, cluster gas mass $\mgas$, and cluster gas mass profiles in logarithmically spaced bins out to $1.5\,R_{500}$, $M_\mathrm{g}(r)$.  The Test Set was built from the \tng simulation and was intentionally designed to be out of domain, utilizing a higher-resolution simulation with a different feedback model and a noisier mock observation strategy.  This was done to approximate a realistic domain shift experiment where we train on a simulation and test on ``reality.''

We explored a continuum of solutions to managing the subtle domain shift from the \magneticum mock observations to the \tng mock observations:  
\begin{itemize}
    \item \mgas, $T$, and \yx power laws:  these traditional scaling relations are used as a benchmark against which to compare the ML models.
    \item NN:  a standard neural network trained on the Training Data; this is the most straightforward and naive approach to ML regression.
    \item SANN: a neural network trained on the \multimodel Training Data, a data set with engineered scatter intended to capture a larger range of physical models and mitigate domain shift.
    \item DRRN:  a Deep Recursive Regression Network, a model designed to force the labeled and unlabeled data to share a latent space, potentially mitigating bias due to domain shift. 
\end{itemize}

On the Training and Validation Sets, the NN appears to be an improved method, outperforming the \yx scaling on the Training Set by 17\%.  However, the model does not transfer to the Test Set, and the model underperforms against the more traditional \yx approach, performing 40\% \textit{worse} than the \yx scaling in terms of scatter, and is furthermore biased toward underprediction, with a median mass error of $-0.02$ dex.

Surprisingly, the SANN model did not outperform the NN model in terms of bias; we expected that training on a noisier data set would mitigate bias due to domain shift, but this was not the case.  A saliency analysis showed that both the NN and SANN learned to excise cluster cores, giving more weight to the cluster gas mass near $R_\mathrm{500}$ and beyond.

Though the DRRN did not result in any remarkable improvement of error or bias, it did have the benefit of ``truth in advertising,'' giving similar results for the Validation and Test Sets.  A t-SNE analysis of the latent space showed that the model was successful in mapping the labeled Training Data and the unlabeled Test Data onto the same latent space.  Domain adaptation in regression tasks is inherently more difficult than in classification. While the model succeeded in learning a shared latent space between the two disjoint data sets, it failed to meet the scientific objective of providing unbiased mass estimates.

How should we interpret these negative results?  Data-Driven Astronomy's goal should not embrace machine learning methods at the expense of science.  We should aim to advance physical understanding by pairing observational data with appropriate methodological solutions, regardless of whether they are exciting, flexible, and modern deep learning tools or rigorous, transparent, and theory-driven statistical approaches.

The analysis presented here raises the question:  Why not simply calibrate the ML mass estimates?  The concept of calibration is firmly established in astronomical practice, and for the scientific question explored here, this is the most logical solution.  Furthermore, it is straightforward enough to calibrate away the offsets shown in Figure \ref{fig:errpdf} with weak lensing masses.  But this reasoning does not extend to deep learning applied to cosmological simulation–based inference.  With only a single Universe against which to test measurements of cosmological parameters, calibration is not a viable option \textemdash{} there is little value in tightening error bars with machine learning if the outputs must ultimately be adjusted to match a predetermined result.

Deep learning should not be treated as a magical black box from which a sufficiently skilled coder can extract a factor-of-ten improvement, but rather as a partner in discovery, warranting careful use and even more careful interpretation. These results remind us that we must approach machine learning tools with rigor and restraint, recognizing that the patterns an algorithm learns from simulations are not necessarily the same ones that are imprinted by nature. \\ 
\bigskip

\acknowledgments{We would like to thank Abigail Crites, Benedikt Diemer, Gus Evrard, Matt Ho, Sarah Loebman, Adam Mantz, Daisuke Nagai, Annalisa Pillepich, Andrea Smith, Krista Lynne Smith, Stephanie Smith, Hy Trac, Jason Tumlinson, Alexey Vikhlinin, and Justin West for their helpful feedback on this project.  

The material presented is based on work supported by NASA under award No. 80NSSC22K0821.  Ana Maria Delgado acknowledges support from NSF grant number 2307070.
Support for JAZ was provided by the {\it Chandra} X-ray Observatory Center, which is operated by the Smithsonian Astrophysical Observatory for and on behalf of NASA under contract NAS8-03060. This work was carried out at the Advanced Research Computing at Hopkins (ARCH) core facility  (rockfish.jhu.edu), which is supported by the National Science Foundation (NSF) grant number OAC 1920103.This work was performed in part at Aspen Center for Physics, which is supported by National Science Foundation grant PHY-2210452. 

This work was produced by Fermi Forward Discovery Group, LLC under Contract No. 89243024CSC000002 with the U.S. Department of Energy, Office of Science, Office of High Energy Physics. The United States Government retains and the publisher, by accepting the work for publication, acknowledges that the United States Government retains a non-exclusive, paid-up, irrevocable, world-wide license to publish or reproduce the published form of this work, or allow others to do so, for United States Government purposes. The Department of Energy will provide public access to these results of federally sponsored research in accordance with the DOE Public Access Plan (\url{http://energy.gov/downloads/doe-public-access-plan}).}

\bibliography{main}
\bibliographystyle{apj}

\end{document}